\documentclass[aip,reprint]{revtex4-2}

\usepackage{amsfonts}
\usepackage{amssymb}
\usepackage{amsmath}
\usepackage[english]{babel}
\usepackage{tumcolors}
\usepackage{tummath}
\usepackage{siunitx}
\usepackage{tikz}
\usepackage[straightvoltages, europeanresistors]{circuitikz}
\usepackage{pgfplots}
\usepackage{pgfplotstable}
\usepackage{environ}
\usepackage[hidelinks]{hyperref}
\usepackage[autostyle, english=american]{csquotes}
\usepackage{orcidlink}
\usepackage{upgreek}
\MakeOuterQuote{"}

\usepackage{multirow}
\usepackage{booktabs}
\usepackage{mathtools}
\usepackage{array}
\usepackage{orcidlink}
\usepackage{listings}

\newcommand{\figref}[2][{}]{\hyperref[#2]{\figurename~\ref{#2}#1}}
\newcommand{\tabref}[2][{}]{\hyperref[#2]{\tablename~\ref{#2}#1}}
\renewcommand{\imu}{\i}
\DeclareMathOperator{\traceu}{Tr}
\DeclareSIUnit{\arbitraryunit}{arb.u.}
\DeclareSIUnit{\period}{period}
\DeclareSIUnit{\dBc}{dBc}

\usetikzlibrary{arrows, decorations.markings, external}
\usetikzlibrary{3d, positioning, fit, calc}
\usetikzlibrary{arrows.meta, shapes, decorations.pathmorphing}

\usepgfplotslibrary{units,fillbetween}
\pgfplotsset{compat=newest}

\pgfplotsset{
  layers/axis lines on top/.define layer set={
      axis grid,
      pre main,
      axis background,
      main,
      axis ticks,
      axis tick labels,
      axis lines,
      axis descriptions,
      axis foreground,
    }{/pgfplots/layers/standard},
}
\pgfplotsset{yaxis color style/.style={y axis line style = {#1},
      y tick label style= {#1},
      y tick style= {#1},
      ylabel style = {#1},
    }}
\newcommand{\rulred}[1]{\textcolor{TUMExtRed}{\underline{\textcolor{black}{#1}}}}
\newcommand{\rulgreen}[1]{\textcolor{TUMGreen}{\underline{\textcolor{black}{#1}}}}

\hypersetup{
  colorlinks=true,
  linkcolor=blue,
  citecolor=blue,
  filecolor=blue,
  urlcolor=blue
}
\begin{document}

\title{Modeling~of~Fluctuations~in~Dynamical~Optoelectronic~Device~Simulations~within~a~Maxwell-Density~Matrix~Langevin~Approach}

\author{Johannes \surname{Popp}\orcidlink{0000-0003-1745-4888}}
\email[]{johannes.popp@tum.de}
\homepage[]{https://www.ee.cit.tum.de/cph/home/}
\author{Johannes \surname{Stowasser}\orcidlink{0009-0007-2240-163X}}
\author{Michael A. \surname{Schreiber}\orcidlink{0000-0002-6003-3897}}
\author{Lukas \surname{Seitner}\orcidlink{0000-0002-0985-8594}}
\affiliation{TUM School of Computation, Information and Technology, Technical
  University of Munich, 85748 Garching, Germany}
\author{Felix \surname{Hitzelhammer}\orcidlink{0009-0006-5335-7363}}
\affiliation{Institute of Physics, NAWI Graz, University of Graz, Universitätsplatz 5, 8010 Graz, Austria}
\author{Michael \surname{Haider}\orcidlink{0000-0002-5164-432X}}
\affiliation{TUM School of Computation, Information and Technology, Technical
  University of Munich, 85748 Garching, Germany}
\author{Gabriela \surname{Slavcheva}\orcidlink{0000-0001-5474-9808}}
\affiliation{Institute of Physics, NAWI Graz, University of Graz, Universitätsplatz 5, 8010 Graz, Austria}
\affiliation{Quantopticon, 5235 South Harper Court, Chicago, IL 60615 USA}
\author{Christian \surname{Jirauschek}\orcidlink{0000-0003-0785-5530}}
\email[]{jirauschek@tum.de}
\affiliation{TUM School of Computation, Information and Technology, Technical
  University of Munich, 85748 Garching, Germany}
\affiliation{TUM Center for Quantum Engineering (ZQE), 85748 Garching,
  Germany}

\date{\today}

\begin{abstract}
  We present a full-wave Maxwell-density matrix simulation tool including c-number stochastic noise terms for the modeling of the spatiotemporal dynamics in active photonic devices, such as quantum cascade lasers (QCLs) and quantum dot (QD) structures. The coherent light-matter interaction in such devices plays an important role in the generation of frequency combs and other nonlinear and nonclassical optical phenomena. Since the emergence of nonlinear and nonclassical features is directly linked to the noise properties, detailed simulations of the noise characteristics are required for the development of low-noise quantum optoelectronic sources.  Our semiclassical simulation framework is based on the Lindblad equation for the electron dynamics, coupled with Maxwell's equations for the optical propagation in the laser waveguide. Fluctuations arising from interactions of the optical field and quantum system with their reservoirs are treated within the quantum Langevin theory. Here, the fluctuations are included by adding stochastic c-number terms to the Maxwell-density matrix equations. The implementation in the mbsolve dynamic simulation framework is publicly available.
\end{abstract}

\maketitle
\section{\label{sec-introduction}Introduction}  An optical frequency comb (OFC) describes
coherent radiation with a broadband spectrum consisting of discrete,
equidistantly spaced optical lines featuring a stable phase relation with low
phase noise and low mode partition
noise.\cite{diddams20202optical,faist2016quantum, fortier201920} Typically,
such combs are used for measurements of optical frequencies in metrology and
sensing, having revolutionized these fields by providing unprecedented
accuracy and enabling numerous innovative applications.\cite{udem2002optical,
  chang2022integrated} Promising semiconductor lasers (SCLs) for integrated
optical frequency comb technologies in the mid-infrared and terahertz (THz)
regime are quantum dot
(QD),\cite{rafailov2005high,rafailov2007mode,lu2008three,
  liu2008dual,rosales2011inas,
  liu2018fs,norman2018perspective,liu2019high,weber2019threshold,pawlus2018intensity,
  hillbrand2020in} quantum
dash,\cite{merghem2009pulse,rosales2012high,merghem2014stability,sooudi2012optical,Panapakkam2016amplitude,
  kemal2019coherent} interband
cascade\cite{meyer2020the,schwarz2019monolithic,bagheri2018passively,sterczewski2019midir}
and quantum cascade lasers (QCLs).
\cite{consolino2019qcl,faist2016quantum,hugi2012mid,singleton2018evidence,cappelli2019retrieval,riccardi2023terahertz,burghoff2014terahertz}
The active gain medium of the aforementioned low-dimensional SCLs provides a
large third-order nonlinearity, which gives rise to a broadband four-wave
mixing (FWM) process resulting in mode
proliferation.\cite{chang2022integrated, faist2016quantum} A complex
interplay of parametric gain, FWM nonlinearity, chromatic dispersion and
spatial hole burning is essential for frequency comb
formation.\cite{hugi2012mid,gioannini2015time,rosales2012high,
  lu2008three,burghoff2020unraveling,bardella2017self,
  dong2017traveling,opacak2019theory,schwarz2019monolithic} For a better
understanding and for improving the laser performance, noise and linewidth
characteristics of such devices have been extensively studied, both
theoretically and
experimentally.\cite{cappelli2015intrinsic,li2019basic,weber2019threshold,bardella2017self,pawlus2018intensity,duan2018carrier,zhou2020optical,borodkin2023noise,gabbrielli2021mid,mao2020pulse,lu2021inas}
Stable and robust OFC operation requires a narrow beatnote, which is a
measure for the amount of amplitude and phase noise of the comb lines. Noise
accompanying carrier transport and spontaneous emission noise can therefore
have a significant impact on the OFC formation and the performance of SCLs.

Significant research efforts are devoted to the generation and deployment of
non-classical features in optical and electronic
systems.\cite{aichhorn2008quantum,giovannetti2011advances, krantz2019quantum,
  hanai2019non, erhard2020advances,daley2022practical} Recently, intensity
correlations in QCL harmonic frequency combs (HFCs) have been experimentally
investigated to develop a new generation of semiconductor devices for
generating light with non-classical properties.\cite{gabbrielli2022intensity}
Endowing commercial devices with outstanding quantum features would pave the
way to practical and high-performance applications in the field of quantum
networks,\cite{kimble2008quantum,wehner2018quantum} including quantum
computation,\cite{ladd2010quantum,zhong2020quantum} quantum
communication,\cite{gisin2007quantum} quantum
metrology,\cite{Degen2017quantum,polino2020photonic,barbieri2022optical} and
quantum simulation.\cite{bloch2012quantum, georgescu2014quantum} Notably,
photonic systems are quite attractive for the investigation and employment of
non-classical features, such as the generation of so-called quantum
combs,\cite{kues2019quantum,yang2021squeezed,guidry2022quantum} corresponding
to non-classical states of light with multimode squeezed and/or entangled
output. As the emergence of non-linear and non-classical features in SCLs is
directly linked to the noise
properties,\cite{rana2002current,gabbrielli2022intensity} the development of
low-noise SCL sources based on detailed simulations is an important
prerequisite.

For the modeling of the optical dynamics in nano-optoelectronic devices, the
Maxwell-Bloch equations are widely used since they form a relatively compact
and numerically efficient model, and thus allow for spatiotemporal
simulations of the laser dynamics over many optical
roundtrips.\cite{allen1987optical,hess1996maxwell,jirauschek2019optoelectronic}
Here, the Bloch equations are used for simulating the evolution of the
quantum system and its coherent light-matter interaction with the optical
field in the active medium. Additionally, the optical field propagation is
treated classically within Maxwell's equations, where the coupling with the
quantum system arises from the macroscopic polarization
term.\cite{jirauschek2019optoelectronic} The density matrix formalism can be
extended and adapted by adding further quantized states in addition to the
laser levels, and tunneling between states.

{Within the Maxwell-density matrix equations, the dissipation processes can be modeled on a microscopic level, by including relevant mechanisms such as carrier-carrier and carrier-phonon scattering. Based on the specific Hamiltonian, relaxation and dephasing rates can be derived within the second-order Born-Markov approximation.\cite{hess1996maxwell,rossi2002theory,iotti2005microscopic,weber2009density} Typically, the dissipation in the Maxwell-density matrix equations is treated by phenomenological dephasing rates and electron lifetimes, which are either empirically chosen or estimated based on experimental data. Alternatively, advanced carrier transport models of different levels of complexity and numerical efficiency can be used to determine the scattering and dephasing rates self-consistently. The ensemble Monte Carlo (EMC) method, for example, is a widely used statistical method for the numerical evaluation of the carrier transport in SCLs, such as QCLs.\cite{kuhn1992monte,jirauschek2014modeling,borowik2017monte,jirauschek2017self} Since the semiclassical EMC model cannot cover quantum coherence effects, such as resonant tunneling across thick injection barriers, e.g., in THz QCLs, quantum corrections based on the density-matrix formalism have been incorporated.\cite{jirauschek2017density, callebaut2005importance} A more general quantum transport approach, which has been applied to the simulation of QCLs, is given by non-equilibrium Green's functions (NEGFs).\cite{lee2002nonequilibrium,wacker2013nonequilibrium,jirauschek2014modeling,winge2016simple,schmielau2009nonequilibrium} Both are self-consistent 3D approaches not relying on empirical or fitting parameters as input and taking into account intrasubband processes between different kinetic energy states within a subband. In general, the NEGF approach contains the full quantum information of scattering and dephasing effects and is therefore superior to semiclassical methods such as the EMC method. However, the numerical simulation of the NEGF solution is expensive, if incoherent scattering within the self-consistent Born approximation is taken into account. To reduce the numerical costs, NEGF simulations are typically executed by neglecting electron-electron scattering beyond the Hartree approximation.\cite{jirauschek2014modeling}}

Typically, the Maxwell-density matrix equations are treated in the so-called
rotating-wave approximation where a coarser spatiotemporal grid can be used
and thus the numerical load is greatly reduced. This approximation is however
only valid for relatively narrowband (and not too strong) optical
fields.\cite{allen1987optical,jirauschek2019optoelectronic,
  slavcheva2002coupled} On the other hand, for frequency comb sources a
broadband spectrum with a high number of modes is desired, and the QCL even
offers the potential for generating spectra extending over a full octave and
beyond.\cite{rosch2015octave} In several studies, dynamical simulations of
different nonlinear optical phenomena in nanostructured devices, e.g. QCLs
\cite{menyuk2009self,talukder2014quantum, tzenov2016time,
  Wang2018rate,Columbo2018dynamics,wang2018harmonic, opacak2019theory,
  silvestri2022multimode, li2019basic, li2022real, rossetti2011time} and QD
lasers,\cite{slavcheva2008model, majer2010cascading, cartar2017self,
  slavcheva2019ultrafast, bardella2017self} were conducted. Motivated by these
developments, over the recent years our group has developed the open-source
codebase mbsolve for the numerically efficient simulation of full-wave
Maxwell-density matrix-type equations, i.e., without invoking the
rotating-wave
approximation.\cite{riesch2021mbsolve,seitner2021group,popp2023self}

\begin{figure}
  \centering
  \resizebox{\columnwidth}{!}{
    \includegraphics{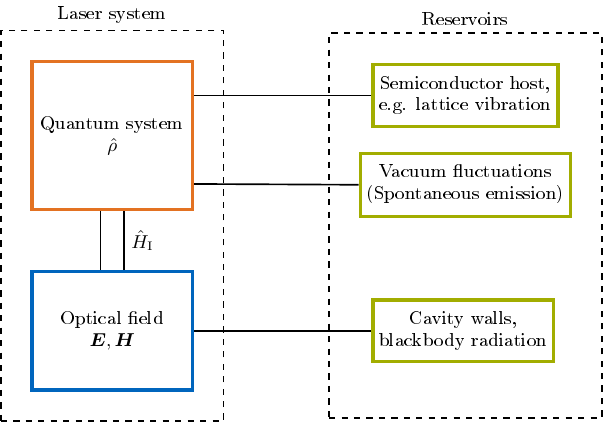}}
  \caption{Schematic illustration of the coupling of a SCL quantum system and field system and the interaction with their associated reservoirs.}
  \label{fig:diag_langevin}
\end{figure}

As pointed out above, this work aims to implement noise sources into mbsolve,
enabling a realistic simulation of the noise characteristics in
low-dimensional SCLs, which is for example important in the context of OFC
operation. In a semiclassical framework, noise can generally be included by
adding stochastic
terms.\cite{polder1979superfluorescence,wodkiewicz1979stochastic} These are
typically numerically implemented by using a pseudorandom number generator
producing uncorrelated, Gaussian-distributed random numbers for every
gridpoint.\cite{slavcheva2004fdtd,andreasen2009finite} The resulting
Maxwell-Bloch equations are then commonly solved numerically with the
finite-difference time-domain
approach.\cite{slavcheva2004fdtd,andreasen2009finite} The magnitude of the
stochastic terms can be systematically derived from the quantum Langevin
equations,\cite{sargent1984laser,gardiner2004quantum,haken1984laser,haken1966quantum}
which are then represented by equivalent stochastic c-number
equations,\cite{slavcheva2004fdtd,graham1968quantum,lax1969quantum,drummond1991quantum}
i.e., evolution equations for operator expectation values with additional
stochastic terms. The c-number Langevin equations have also been used to
calculate the intrinsic linewidth and estimate the intensity noise in
SCLs.\cite{haug1967theory,liu2012effects,Wang2018rate,essebe2019stochastic,rana2002current,gensty2005semiclassical,duan2018carrier,zhou2020optical}
Spontaneous emission obviously plays an important role in optoelectronic
devices. While the resulting recombination can simply be included by
nonlinear rate terms for the carrier
occupations,\cite{majer2010cascading,gehrig2002mesoscopic} the noise
contribution is not included in the Maxwell-Bloch model due to its
semiclassical nature. This effect can however be considered in terms of a
Gaussian white noise source in the optical propagation
equation.\cite{slavcheva2004fdtd, kim2010maxwell,wilkinson2012influence} In a
different model, dipole fluctuations are also included by adding Langevin
noise terms not only to the propagation equation, but also to the
off-diagonal density matrix
elements.\cite{gehrig2001spatio,gehrig2002mesoscopic,gehrig2003,hoffmann1999quantum,
  Cerjan2015quantitative, benediktovitch2019quantum} By virtue of the
fluctuation-dissipation theorem, a decay of populations, coherences, or the
optical field is generally accompanied by fluctuations, and a Maxwell-Bloch
equation model which includes such decay-induced fluctuations has been
presented.\cite{andreasen2009finite,pusch2012coherent,fang2021modelling} We
extend this approach by including incoherent tunneling. From this, we can
derive the semiclassical noise terms for our Maxwell-density matrix Langevin
approach, which we then incorporate in our open-source tool mbsolve. Notably,
an extension of the stochastic c-number approach to incorporate nonclassical
effects has been discussed,\cite{drummond1981quantum,drummond1991quantum}
which might enable a direct investigation of nonclassical light generation in
the presence of noise.

In detail, the paper is organized as follows: In Section~\ref{sec:model}, we
calculate the stochastic noise terms of the quantum Langevin equations and
derive the generalized description of the Maxwell-density matrix Langevin
equations for a quantum-optoelectronic structure such as a QCL. Our model is
illustrated schematically in Fig.~\ref{fig:diag_langevin}. Here, the
structure is described in terms of the optical field represented by the
electric and magnetic field vectors $\vec{E}\,,\vec{H}$ and the density
operator $\op{\rho}$, which are coupled to each other by the interaction
Hamiltonian $\op{H}_\mathrm{I}$. For the calculation of drift and diffusion
operators in the quantum Langevin theory, we take into account the influence
of various reservoirs in our structure. Regarding the quantum system, the
reservoir interactions with the semiconductor host, which for example
includes phonons associated with (longitudinal- and transverse-optical and
-acoustic) thermal lattice vibrations, lattice imperfections in the form of
impurities (such as dopants), interface roughness or atomic disorder in
alloys, as well as vacuum fluctuations arising from spontaneous emission are
considered. For the optical field, the interaction with noise arising from
thermal radiation (blackbody) entering the active waveguide from the cavity
walls can be taken into account by external noise
sources.\cite{andreasen2008finite} In this paper, we rather focus on the
fluctuations arising from the quantum system and dedicate the investigation
of thermal noise influences, which can for example play a role in THz QCL
active waveguides, to future works. In Section~\ref{sec:simulation}, we give
an overview of the simulation tool and the implementation of the noise terms
and validate the model by presenting the simulation results for a
superfluorescence setup.\cite{andreasen2009finite,maki1989influence}
Furthermore, we present the simulation results for a THz QCL harmonic
frequency comb and discuss the effects of noise contributions on the comb
characteristics. The paper is concluded in Section~\ref{sec:conclusion}.
\section{\label{sec:model}Theoretical model}
{In the following, we restrict ourselves to optoelectronic devices where the optical field can be well modeled using 1D Maxwell's equations. This for example applies to semiconductor lasers with longitudinally invariant waveguide geometries, where the 3D Maxwell's equations can be reduced to an effective 1D model.\cite{jirauschek2019optoelectronic} The focus lies on the inclusion of noise arising for example from spontaneous emission and fluctuations associated with the electron transport.} First, we introduce the quantum Langevin equations using a simple three-level resonant tunneling QCL system. {Here, the reservoir variables are eliminated and replaced by drift and fluctuation terms within the Heisenberg equation of motion.} The quantum Langevin equations can be transformed into associated c-number Langevin equations, which are then used to derive the stochastic noise terms incorporated into the full-wave Maxwell-density matrix equation system.

\subsection{The quantum Langevin equations}
\begin{figure}
  \centering
  \includegraphics{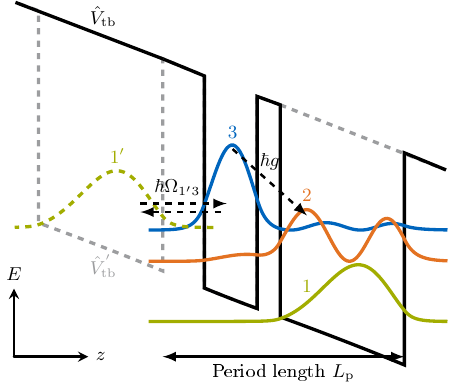}
  \caption{Schematic conduction band profile and probability densities of a two well THz QCL structure, where the upper laser level $3$ is populated via resonant tunneling from injector level $1'$. Depopulation occurs through LO-phonon scattering from the lower laser level $2$ to the depopulation level $1$. }
  \label{fig:wf_two_well}
\end{figure}

The quantum Langevin equations are introduced by using a simple three-level
resonant tunneling QCL system as depicted in
Fig.~\ref{fig:wf_two_well}.\cite{kumar2009coherence, tzenov2016time} The QCL
exploits optical transitions between quantized states in the conduction band
of a quantum well heterostructure, where the properties can be controlled by
quantum design rather than being determined by the bulk material. This not
only applies to the gain and lasing wavelength, but also to the nonlinear
optical properties such as FWM. Besides confinement provided by the quantum
wells, another important quantum effect is tunneling through the separating
barriers, which significantly influences carrier transport, in addition to
the incoherent scattering-induced transitions due to phonons, crystal
imperfections and electron-electron
interactions.\cite{jirauschek2014modeling,jirauschek2017density} Regarding
non-stationary QCL operation as is the case for OFC emission, coherent
light-matter interaction as a further quantum effect plays a significant role
in the dynamic behavior, e.g., leading to Rabi
flopping,\cite{wang2007coherent} i.e. oscillations of the electron population
between the upper and lower laser levels driven by the resonant optical
field. Dephasing due to incoherent scattering has to be taken into account
for a realistic description, as it greatly affects tunneling and coherent
light-matter interaction.

For the structure shown in Fig.~\ref{fig:wf_two_well}, the lasing transition
occurs between the upper laser level $|3\rangle$ and the lower laser level
$|2\rangle$. Depopulation takes place via level $|1\rangle$ and electrons are
injected from the depopulation level $|1'\rangle$ of the adjacent period via
resonant tunneling. The resonant tunneling across thick injection barriers in
THz QCLs is treated within the tight-binding
model.\cite{callebaut2005importance,kumar2009coherence,terazzi2010density,jirauschek2017self,jirauschek2014modeling,dupont2010simplified}
Here, the tunneling between a doublet of states at the thick injection
barrier is described by the coupling strength $\Omega_{ij} =
  -\hbar^{-1}\langle i |\op{V}_\mathrm{ext} - \op{V}_\mathrm{tb}| j \rangle$,
with the extended conduction band potential $\op{V}_\mathrm{ext}$ and the
tight-binding potential $\op{V}_\mathrm{tb}$. The coupling strengths
$\Omega_{ij}$ between the states $|3\rangle,\, |2\rangle,\, |1\rangle$ within
the active period are zero.\cite{terazzi2010density}

In general, the QCL laser system is then described by the reduced system
Hamiltonian\cite{scully1997quantum,louisell1973quantum,liu2012effects}
\begin{equation}
  \begin{split}
    \op{H_\mathrm{s}} ={}& \op{H}_\mathrm{F} + \op{H}_0 + \Delta\op{V}_\mathrm{tb}+ \op{H}_\mathrm{I} \\
    ={}& \hbar\omega_0 \op{a}^\dagger\op{a} + \sum_i \epsilon_i |i\rangle \langle i| - \hbar \Omega_{1'3} \left( |1'\rangle \langle 3| + |3\rangle \langle 1'| \right) \\
    {} & + \hbar g \left(|3\rangle \langle 2| + |2\rangle \langle 3|\right) \left(\op{a} +\op{a}^\dagger \right) \,,
  \end{split}
  \label{eq:hamil_sys}
\end{equation}
where $\op{H}_\mathrm{F}$ is the Hamiltonian of the optical field, $\op{H}_0$ is the Hamiltonian of the quantum system with $\Delta \op{V}_\mathrm{tb}$ describing the coupling of electron states in two adjacent periods within the tight-binding model, and $\op{H}_\mathrm{I}$ constitutes the interaction Hamiltonian between quantum system and optical field.  Here, $\hbar$ is the reduced Planck constant, $\omega_0$ the single mode lasing angular frequency, $\op{a}^\dagger(\op{a})$ denotes the creation (annihilation) operator of the radiation field, $\epsilon_i$ is the energy of level $|i\rangle$ and $\hbar\Omega_{1'3}$ the anticrossing energy gap between levels $|1'\rangle$ and $|3\rangle$. The dipole coupling constant $g$ can be written in terms of the dipole matrix element, $\mu_{z,23}= q \langle 2 | \op{z} | 3\rangle$, as\cite{scully1997quantum,sargent1984laser}
\begin{equation}
  g = -\sqrt{\frac{\omega_0}{2 \hbar \epsilon_r \epsilon_0 V_\mathrm{p}}} \mu_{z,23}\,,
\end{equation}
where
$\epsilon_\mathrm{r}$ is the relative permittivity, $\epsilon_0$ is the vacuum permittivity and $V_\mathrm{p}$ is the volume of each quantum system associated with an active QCL period.

The Heisenberg-Langevin equation of motion for an operator
$\function{\op{A}_\mu}{t}$ reads as\cite{haken1984laser,louisell1973quantum,
  lax1966quantum,sargent1984laser}
\begin{equation}
  \begin{split}
    \partial_t{\function{\op{A}_\mu}{t}}={}&
    -\imu\hbar^{-1}\commutator{\function{\op{A}_\mu}{t}}{\function{\op{H}_\mathrm{s}}{t}}+\function{\op{D}_\mu}{t}+\function{\op{F}_\mu}{t}\\
    ={}& \function{\op{M}_\mu}{t}+\function{\op{F}_\mu}{t} \,.
  \end{split}
  \label{eq:heis-lang}
\end{equation}
Here, the drift operator $\function{\op{D}_\mu}{t}$ and fluctuation operator
$\function{\op{F}_\mu}{t}$ account for the influence of the reservoirs on the
system. $\commutator{\cdot}{\cdot}$ denotes the commutator
$\commutator{\op{X}}{\op{Y}}=\op{X}\op{Y}-\op{Y}\op{X}$. {For the drift
operator $\op{D}_\mu$ we can under the Markovian approximation
write\cite{lax1966quantum,louisell1973quantum}
\begin{equation}
  \begin{split}
    {\op{D}_\mu} ={}& - \sum_{i,j} \function{\delta}{\omega_i, -\omega_j} \left\{ \commutator{\op{A}_\mu}{\op{Q}_i} \op{Q}_j w_{ij}^+ \right.\\ {}& \left. - \op{Q}_j \commutator{\op{A}_\mu}{\op{Q}_i} w_{ji}^- \right\}\,,
  \end{split}
  \label{eq:diff-calc}
\end{equation}
where $w^\pm$ are the reservoir spectral densities and $\op{Q}_i$ is a function of system operators. For a detailed description and derivation of this theory together with the calculation examples for specific operators $\op{A}_\mu$, we refer to Refs.~\onlinecite{lax1966quantum, louisell1973quantum}.}

The reservoir average of the fluctuation operator vanishes,
$\langle\op{F}^\dagger_{\mu}\rangle_\mathrm{R} =
  \langle\op{F}_{\mu}\rangle_\mathrm{R} = 0$. {The diffusion coefficient for a
Markovian system is defined as}
\begin{equation}
  \function{2\langle \op{D}_{\mu \nu}}{t}\rangle_\mathrm{R} \function{\delta} {t -t'} = \langle \function{\op{F}_{\mu}}{t} \function{\op{F}_{\nu}}{t'}\rangle_\mathrm{R}\,,
\end{equation}
and can be calculated by applying the fluctuation-dissipation theorem. {Here,
  the $\delta$-function indicates the very short memory period of the
  reservoirs.} The \textit{generalized Einstein relation} for the calculation
of the diffusion coefficient is given
by\cite{lax1966quantum,marcuse1984computer,sargent1984laser}
\begin{equation}
  \begin{split}
    \function{2\langle \op{D}_{\mu \nu}}{t}\rangle_\mathrm{R} = {}& \partial_t \langle \function{\op{A}_\mu}{t}
    \function{\op{A}_\nu}{t}\rangle_\mathrm{R} -
    \langle\function{\op{M}_\mu}{t}\function{\op{A}_\nu}{t}\rangle_\mathrm{R} \\
    {} & -
    \langle\function{\op{A}_\mu}{t}\function{\op{M}_\nu}{t}\rangle_\mathrm{R}\,.
    \label{eq:einstein}
  \end{split}
\end{equation}
From Eq.~\eqref{eq:heis-lang} together with Eqs.~\eqref{eq:hamil_sys} and
\eqref{eq:diff-calc} the quantum Langevin equations for the three-level QCL
quantum system can be derived. Therefore, we introduce the electron
population operators $\op{\sigma}_{ii} = |i\rangle\langle i |$ and the
coherence operators $\op{\sigma}_{ij} = |i\rangle\langle j |$. The term
$\op{\sigma}_{32}\op{a}^\dagger$ describes the creation of a photon
accompanied by an electron transition from the lower to the higher lying
energy level, and $\op{\sigma}_{23}\op{a}$ the annihilation of a photon
accompanied by an electron transition from the higher to the lower lying
energy level. {At this point we drop these counterrotating energy
  non-conserving terms in the interaction Hamiltonian $\op{H}_\mathrm{I}$ as in
  the commonly used rotating wave approximation.\cite{sargent1984laser,
    scully1997quantum} This simplifies the following calculations of the noise
  terms. A more complete calculation should also include more than one mode of
  the optical field in the system Hamiltonian. However, these concessions do
  not affect the form of the specific noise terms which are ultimately used in
  our simulations.} The corresponding equations of motion are given by
\begingroup
\allowdisplaybreaks
\begin{subequations}
  \begin{align}
    \partial_t \function{\op{a}}{t} ={}             & -\imu \omega_0 \function{\op{a}}{t}
    -\frac{\kappa}{2} \function{\op{a}}{t} - g \op{\sigma}_{23} +
    \function{\op{F}_{a}}{t}\,,                                                                                 \\
    \partial_t \function{\op{\sigma}_{23}}{t} ={}   &
    -\frac{\imu}{\hbar}\Delta_{32}\function{\op{\sigma}_{23}}{t}
    -\gamma_{23}\function{\op{\sigma}_{23}}{t} + \imu \Omega_{1'3}
    \function{\op{\sigma}_{21'}}{t} \nonumber                                                                   \\ {} & + \imu g
       \left[\function{\op{\sigma}_{33}}{t} - \function{\op{\sigma}_{22}}{t}\right]
    \function{\op{a}}{t} + \function{\op{F}_{23}}{t}\,,
    \\ \partial_t \function{\op{\sigma}_{31'}}{t} ={} & -\frac{\imu}{\hbar}\Delta_{1'3}
    \function{\op{\sigma}_{31'}}{t} -\gamma_{1'3} \function{\op{\sigma}_{31'}}{t}
    + \imu \Omega_{1'3} \left[\function{\op{\sigma}_{33}}{t} \right. \nonumber                                  \\
    {}                                              & \left. -\function{\op{\sigma}_{1'1'}}{t} \right] + \imu g
    \function{\op{\sigma}_{21'}}{t} \function{\op{a}^\dagger}{t} +
    \function{\op{F}_{31'}}{t}\,,                                                                               \\
    \partial_t \function{\op{\sigma}_{21'}}{t} ={}  & -\frac{\imu}{\hbar}\Delta_{1'2}
    \function{\op{\sigma}_{21'}}{t} -\gamma_{1'2} \function{\op{\sigma}_{21'}}{t}
    + \imu \Omega_{1'3} \function{\op{\sigma}_{23}}{t} \nonumber                                                \\ {} & + \imu g
       \function{\op{\sigma}_{31'}}{t} \function{\op{a}}{t} +
    \function{\op{F}_{21'}}{t}\,,                                                                               \\
    \partial_t \function{\op{\sigma}_{33}}{t} ={}   & - \frac{1}{\tau_3}
    \function{\op{\sigma}_{33}}{t} + r_{32}\function{\op{\sigma}_{22}}{t} +
    r_{31'}\function{\op{\sigma}_{1'1'}}{t} \nonumber                                                           \\ {} & + \imu g \left[
         \function{\op{a}^\dagger}{t}\function{\op{\sigma}_{23}}{t} -
         \function{\op{a}}{t}\function{\op{\sigma}_{23}^\dagger}{t}\right] \nonumber
    \\  & - \imu \Omega_{1'3} \left[ \function{\op{\sigma}_{31'}^\dagger}{t} -
    \function{\op{\sigma}_{31'}}{t}\right] + \function{\op{F}_{33}}{t}\,,                                       \\
    \partial_t \function{\op{\sigma}_{22}}{t} ={}   &
    r_{23}\function{\op{\sigma}_{33}}{t} - \frac{1}{\tau_2}
    \function{\op{\sigma}_{22}}{t} + r_{21'}\function{\op{\sigma}_{1'1'}}{t}
    \nonumber                                                                                                   \\  & + \imu g\left[
         \function{\op{a}}{t}\function{\op{\sigma}_{23}^\dagger}{t} -
         \function{\op{a}^\dagger}{t}\function{\op{\sigma}_{23}}{t}\right] +
    \function{\op{F}_{22}}{t}\,,                                                                                \\
    \partial_t \function{\op{\sigma}_{1'1'}}{t} ={} &
    r_{1'3}\function{\op{\sigma}_{33}}{t} + r_{1'2}\function{\op{\sigma}_{22}}{t}
    - \frac{1}{\tau_{1'}} \function{\op{\sigma}_{1'1'}}{t} \nonumber                                            \\ {} & -
       \imu \Omega_{1'3} \left[\function{\op{\sigma}_{31'}}{t} -
         \function{\op{\sigma}_{31'}^\dagger}{t}\right] +
       \function{\op{F}_{1'1'}}{t}\,,
  \end{align}\label{eq:Heisenberg}%
\end{subequations}
\endgroup
where $\kappa$ is the cavity decay rate, $\Delta_{ij}$ denotes the energy
separation between levels $|i\rangle$ and $| j \rangle$, $\tau_i^{-1} =
  \sum_{i \neq j} r_{ji}$ is the inverse population lifetime, $r_{ij, i\neq j}$
represents the scattering rate from level $j$ to $i$ and
\begin{equation}
  \gamma_{ij} = \frac{1}{2} \left( \frac{1}{\tau_i} + \frac{1}{\tau_j}\right) + \gamma_{ij,\mathrm{p}}
\end{equation}
is the dephasing rate. Here, $\gamma_{ij,\mathrm{p}}$  is the pure dephasing rate, which for QCLs mainly consists of elastic scattering contributions due to impurity and interface roughness.\cite{jirauschek2017density} {The equivalent equations for $\function{\op{a}^\dagger}{t}$, $\function{\op{\sigma}_{32}}{t}$, $\function{\op{\sigma}_{1'3}}{t}$ and $\function{\op{\sigma}_{1'2}}{t}$ are given by the Hermitian conjugates of Eqs.~\eqref{eq:Heisenberg}(a)-(d).}

Using Eq.~\eqref{eq:einstein}, we can calculate the second-order correlation
function relevant for the polarization operator as
\begin{equation}
  \label{eq:d3223}
  \begin{split}
    \langle \function{\op{F}_{23}^\dagger}{t} \function{\op{F}_{23}}{t'}\rangle_\mathrm{R} ={} &  2\langle \function{\op{D}_{3223}}{t}\rangle_\mathrm{R} \function{\delta} {t -t'} \\ ={} & \left[ \partial_t \langle \function{\op{\sigma}_{23}^\dagger}{t}
    \function{\op{\sigma}_{23}}{t} \rangle_\mathrm{R} - \langle
    \function{\op{M}_{23}^\dagger}{t} \function{\op{\sigma}_{23}}{t}
    \rangle_\mathrm{R} \right. \\ {}& \left. - \langle
    \function{\op{\sigma}_{23}^\dagger}{t} \function{\op{M}_{23}}{t}
    \rangle_\mathrm{R} \right] \function{\delta} {t-t'}\\ ={} &
    \bigg[\left(2\gamma_{23}
    -\frac{1}{\tau_3}\right)\langle\function{\op{\sigma}_{33}}{t}\rangle_\mathrm{R}
    \\ {} & + r_{32} \langle\function{\op{\sigma}_{22}}{t}\rangle_\mathrm{R} +
    r_{31'}\langle\function{\op{\sigma}_{1'1'}}{t}\rangle_\mathrm{R}\bigg] \\ {}&
    \times \function{\delta} {t -t'} \,.
  \end{split}
\end{equation}
A detailed derivation of the diffusion coefficient $2\langle \function{\op{D}_{3223}}{t}\rangle_\mathrm{R}$ can be found in Appendix~\ref{sec:appendix-diff-coeff}.

With the same procedure, we determine the other non-vanishing second-order
correlation functions\cite{lax1966quantum,louisell1973quantum}
\begingroup
\allowdisplaybreaks
\begin{subequations}
  \begin{align}
    \langle \function{\op{F}_{a}^\dagger}{t} \function{\op{F}_{a}}{t'}\rangle_\mathrm{R} ={}     & \kappa \function{n_\mathrm{th}}{\omega_0} \function{\delta}{t-t'}\,,                                                                                           \\
    \langle \function{\op{F}_{a}}{t} \function{\op{F}_{a}^\dagger}{t'}\rangle_\mathrm{R} ={}     & \kappa \left(\function{n_\mathrm{th}}{\omega_0} + 1 \right) \function{\delta}{t-t'}\,,                                                                         \\ \label{subeq:dipole_noise}
    \langle \function{\op{F}_{23}}{t} \function{\op{F}_{23}^\dagger}{t'}\rangle_\mathrm{R} ={}   & \bigg[r_{23} \langle\function{\op{\sigma}_{33}}{t}\rangle_\mathrm{R} + \left( 2 \gamma_{23} - \frac{1}{\tau_2} \right) \nonumber                               \\ {} & \times \langle\function{\op{\sigma}_{22}}{t}\rangle_\mathrm{R} + r_{21'}\langle\function{\op{\sigma}_{1'1'}}{t}\rangle_\mathrm{R}\bigg] \nonumber                                                         \\ {} & \function{\delta}{t-t'}\,, \\
    \langle \function{\op{F}_{31'}^\dagger}{t} \function{\op{F}_{31'}}{t'}\rangle_\mathrm{R} ={} & \bigg[r_{1'3} \langle\function{\op{\sigma}_{33}}{t}\rangle_\mathrm{R} + r_{1'2} \langle\function{\op{\sigma}_{22}}{t}\rangle_\mathrm{R} \nonumber              \\ {} & + \left(2\gamma_{1'3} - \frac{1}{\tau_{1'}}\right) \langle\function{\op{\sigma}_{1'1'}}{t}\rangle_\mathrm{R}\bigg] \nonumber \\ {}& \times \function{\delta}{t-t'}\,,                                     \\
    \langle \function{\op{F}_{31'}}{t} \function{\op{F}_{31'}^\dagger}{t'}\rangle_\mathrm{R} ={} & \bigg[\left(2\gamma_{1'3} -\frac{1}{\tau_3}\right)\langle\function{\op{\sigma}_{33}}{t}\rangle_\mathrm{R} \nonumber                                            \\ {} & + r_{32} \langle\function{\op{\sigma}_{22}}{t}\rangle_\mathrm{R} + r_{31'} \function{\op{\sigma}_{1'1'}}{t} \bigg] \nonumber                                                         \\ {} & \times \function{\delta}{t-t'} \,,                                                                  \\
    \langle \function{\op{F}_{21'}^\dagger}{t} \function{\op{F}_{21'}}{t'}\rangle_\mathrm{R} ={} & \bigg[r_{1'3} \langle\function{\op{\sigma}_{33}}{t}\rangle_\mathrm{R} + r_{1'2} \langle\function{\op{\sigma}_{22}}{t}\rangle_\mathrm{R} \nonumber              \\ {} &+ \left(2\gamma_{1'2} - \frac{1}{\tau_{1'}}\right) \langle\function{\op{\sigma}_{1'1'}}{t}\rangle_\mathrm{R}\bigg] \nonumber \\ {} & \times \function{\delta}{t-t'}\,,                                     \\
    \langle \function{\op{F}_{21'}}{t} \function{\op{F}_{21'}^\dagger}{t'}\rangle_\mathrm{R} ={} & \bigg[r_{23}\langle\function{\op{\sigma}_{33}}{t}\rangle_\mathrm{R} + \left( 2 \gamma_{1'2} - \frac{1}{\tau_2} \right) \nonumber                               \\ {} &\times \langle\function{\op{\sigma}_{22}}{t}\rangle_\mathrm{R} + r_{21'}\langle\function{\op{\sigma}_{1'1'}}{t}\rangle_\mathrm{R}\bigg] \nonumber \\{} & \times \function{\delta}{t-t'}\,, \\
    \langle \function{\op{F}_{33}}{t} \function{\op{F}_{33}}{t'}\rangle_\mathrm{R} ={}           & \bigg[ \frac{1}{\tau_3} \langle \function{\op{\sigma}_{33}}{t} \rangle_\mathrm{R} + r_{32} \langle \function{\op{\sigma}_{22}}{t} \rangle_\mathrm{R} \nonumber \\ {} & + r_{31'} \langle \function{\op{\sigma}_{1'1'}}{t} \rangle_\mathrm{R} \bigg] \function{\delta}{t-t'}\,,                                                             \\
    \langle \function{\op{F}_{22}}{t} \function{\op{F}_{22}}{t'}\rangle_\mathrm{R} ={}           & \bigg[ r_{23} \langle \function{\op{\sigma}_{33}}{t} \rangle_\mathrm{R} + \frac{1}{\tau_2} \langle \function{\op{\sigma}_{22}}{t} \rangle_\mathrm{R}\nonumber  \\ {}& + r_{21'} \langle \function{\op{\sigma}_{1'1'}}{t} \rangle_\mathrm{R} \bigg]\function{\delta}{t-t'}\,,      \\
    \langle \function{\op{F}_{1'1'}}{t} \function{\op{F}_{1'1'}}{t'}\rangle_\mathrm{R} ={}       & \bigg[ r_{1'3} \langle \function{\op{\sigma}_{33}}{t} \rangle_\mathrm{R} +  r_{1'2} \langle \function{\op{\sigma}_{22}}{t} \rangle_\mathrm{R} \nonumber        \\ {}& + \frac{1}{\tau_{1'}} \langle \function{\op{\sigma}_{1'1'}}{t} \rangle_\mathrm{R} \bigg]\function{\delta}{t-t'}\,.
  \end{align}
\end{subequations}
\endgroup
Here, $\function{n_{\mathrm{th}}}{\omega_0}= \left[\exp \left(\frac{\hbar
      \omega_0}{k_\mathrm{B}T} \right) - 1\right]^{-1}$ is the number of thermal
photons in the lasing mode at temperature $T$, where $k_\mathrm{B}$ denotes
the Boltzmann constant.
\subsection{The c-number Langevin equations}
\label{subsec:c-numberLangevin}
In order to derive the stochastic noise terms for the semiclassical
Maxwell-density matrix equations, the operator Langevin equations have to be
converted into the associated c-number Langevin equations.

The quantum Langevin equation for the operator $\function{\op{A}_{\mu}}{t}$
in chosen order is given by
\begin{equation}
  \partial_t \function{\op{A}_{\mu}}{t} =
  -\imu\hbar^{-1}\commutator{\function{\op{A}_\mu}{t}}{\function{\op{H}_\mathrm{s}}{t}}^\mathrm{c}
  + \function{\op{D}_\mu^\mathrm{c}}{t} + \function{\op{F}_\mu^\mathrm{c}}{t}
  \,,
\end{equation}
where we make use of the commutation relation $\op{A}_\mu^\dagger\op{A}_\nu = \op{A}_\nu \op{A}_\mu^\dagger - \commutator{\op{A}_\nu}{\op{A}_\mu^\dagger}$ to bring the equation into the chosen order.  {We use the superscript $c$ to highlight that we have put the operators in chosen order. To explain this formulation in more detail, we use the fluctuation operator $\function{\op{F}_\mu}{t}$ as an example, but the following description holds for the other operators in the same way. For a chosen order $\op{A}_1, \ldots, \op{A}_\mu$, we can write
  \begin{equation}
    \op{F}_\mu = \function{\op{F}_\mu^\mathrm{c}}{\op{A}_1, \ldots, \op{A}_\mu}\,,
  \end{equation}
  where the fluctuation operator $\op{F}_\mu^\mathrm{c}$ in the chosen order is, of course, equal to the fluctuation operator $\op{F}_\mu$ in the original order.
  The associated c-number fluctuation term $\function{F_\mu^c}{A_1,\ldots,A_\mu}$ is obtained using c-numbers $A_\nu$. By defining a linear chosen ordering operator $\op{C}$ we can further indicate\cite{lax1968quantum,louisell1973quantum}
  \begin{equation}
    \function{\op{F}_\mu^\mathrm{c}}{\op{A}_1, \ldots, \op{A}_\mu} = \function{\op{C}}{\function{F_\mu^c}{A_1,\ldots,A_\mu}}\,,
  \end{equation}
  where the operator $\op{C}$ has the function of replacing each ${A}_\nu$ by the corresponding operator $\op{A}_\nu$ and bringing all terms into chosen order.}

If we now convert the quantum Langevin equation into the equivalent c-number
Langevin equation, we may write
\begin{equation}
  \begin{split}
    \partial_t \function{A_\mu}{t} ={}& \function{L_{\mu}}{t} + \function{D_\mu}{t} +
    \function{F_\mu^\mathrm{c}}{t}\\ ={}& \function{M_\mu}{t} +
    \function{F_\mu^\mathrm{c}}{t}\,,
  \end{split}
  \label{eq:c-numberLangevinEq}
\end{equation}
with $\function{L_\mu}{t}$ being the coherent term corresponding to the commutation of $\function{\op{A}_\mu}{t}$ with the system Hamiltonian $\op{H}_\mathrm{s}$, and $\function{D_\mu}{t}$ denoting the drift term.
Furthermore, by the use of Eq.~\eqref{eq:c-numberLangevinEq} we obtain the c-number equation
\begin{equation}
  \begin{split}
    \partial_t \left[\function{A_\mu}{t} \function{A_\nu}{t} \right]={}&
    \function{A_\mu}{t} \partial_t \function{A_\nu}{t} + \function{A_\nu}{t} \partial_t \function{A_\mu}{t} \\ ={}& \function{A_\mu}{t} \function{M_\nu}{t} +
    \function{A_\nu}{t}\function{M_\mu}{t} \\ {} & + \function{A_\mu}{t}
    \function{F_\nu^c}{t} + \function{A_\nu}{t} \function{F_\mu^c}{t}
    \,.
  \end{split}
  \label{eq:c-numberLangevinEq2}
\end{equation}
In analogy to the reservoir average in the operator case, we may write the c-number equation
\begin{equation}
  \label{eq:cnum_diff}
  \begin{split}
    \partial_t \langle \function{A_\mu}{t} \function{A_\nu}{t} \rangle_\mathrm{R} =
    {}& \langle \function{A_\mu}{t} \function{M_\nu}{t}\rangle_\mathrm{R} +
    \langle\function{A_\nu}{t} \function{M_\mu}{t}\rangle_\mathrm{R}\\ {}& +
    2\langle \function{D_{\mu\nu}}{t} \rangle_\mathrm{R} \,,
  \end{split}
\end{equation}
where we can make use of the following relation under the Markovian approximation:\cite{louisell1973quantum,sargent1984laser}
\begin{equation}
  2\langle \function{D_{\mu\nu}}{t} \rangle_\mathrm{R} = \langle \function{A_\mu}{t} \function{F_\nu^c}{t} + \function{A_\nu}{t} \function{F_\mu^c}{t} \rangle_\mathrm{R} \,.
\end{equation}
The diffusion coefficients in the c-number Langevin equations may differ from the ones in the quantum Langevin equations, as the c-numbers commute, whereas the operators do not. {By requiring the equivalence of Eq.~\eqref{eq:cnum_diff} and Eq.~\eqref{eq:einstein} in both c-number and quantum Langevin theory, it can be shown that in general
  \begin{equation}
    \function{2\langle \op{D}_{\mu \nu}}{t}\rangle_\mathrm{R} \neq 2\langle \function{\op{C}}{\function{D_{\mu\nu}}{t}} \rangle_\mathrm{R}\,.
  \end{equation}}

{By taking our chosen ordered operator representation of the system operators
  $\op{a}^\dagger,\,\op{\sigma}_{23}^\dagger,\,\op{\sigma}_{31'}^\dagger,\,\op{\sigma}_{21'}^\dagger,\,\op{\sigma}_{33},\,\op{\sigma}_{22},\,\op{\sigma}_{1'1'},\,\op{\sigma}_{21'},\,\op{\sigma}_{31'},
    \,\op{\sigma}_{23},\,\op{a}$, we obtain the corresponding c-numbers $a^*,\,
    \sigma_{23}^*,\, \sigma_{31'}^*,\, \sigma_{21'}^*,\, \sigma_{33},\,
    \sigma_{22},\, \sigma_{1'1'},\, \sigma_{21'},\, \sigma_{31'},\,
    \sigma_{23},\,a$.} Therefore, we derive the c-number second-order moments and
obtain for the populations differing terms compared to the operator case:
\begingroup
\allowdisplaybreaks
\begin{widetext}
  \begin{subequations}
    \begin{align}
      \langle \function{F_{33}}{t} \function{F_{33}}{t'}\rangle_\mathrm{R}  = {}   & \bigg\{ \frac{1}{\tau_3} \langle \function{\sigma_{33}}{t} \rangle_\mathrm{R} + r_{32} \langle \function{\sigma_{22}}{t} \rangle_\mathrm{R}  + r_{31'} \langle \function{\sigma_{1'1'}}{t} \rangle_\mathrm{R}  + \imu g \left[ \langle \function{a^*}{t} \function{\sigma_{23}}{t} \rangle_\mathrm{R} - \langle \function{a}{t} \function{\sigma^*_{23}}{t} \rangle_\mathrm{R}\right]  \nonumber                          \\ {} & + \imu \Omega_{1'3} \left[\langle \function{\sigma^*_{31'}}{t} \rangle_\mathrm{R} - \langle\function{\sigma_{31'}}{t}\rangle_\mathrm{R}\right]  \bigg\} \function{\delta}{t-t'}\,,
      \\
      \langle \function{F_{22}}{t} \function{F_{22}}{t'}\rangle_\mathrm{R} ={}     & \left\{  r_{23} \langle \function{\sigma_{33}}{t} \rangle_\mathrm{R} + \frac{1}{\tau_2} \langle \function{\sigma_{22}}{t} \rangle_\mathrm{R} + r_{21'} \langle \function{\sigma_{1'1'}}{t} \rangle_\mathrm{R} + \imu g \left[ \langle \function{a^*}{t} \function{\sigma_{23}}{t} \rangle_\mathrm{R} -\langle \function{a}{t} \function{\sigma^*_{23}}{t} \rangle_\mathrm{R}  \right] \right\} \function{\delta}{t-t'}\,,
      \\
      \langle \function{F_{1'1'}}{t} \function{F_{1'1'}}{t'}\rangle_\mathrm{R} ={} & \left\{ r_{1'3} \langle \function{\sigma_{33}}{t} \rangle_\mathrm{R} +  r_{1'2} \langle \function{\sigma_{22}}{t} \rangle_\mathrm{R} + \frac{1}{\tau_{1'}} \langle \function{\sigma_{1'1'}}{t} \rangle_\mathrm{R} + \imu \Omega_{1'3} \left[ \function{ \langle\sigma^*_{31'}}{t} \rangle_\mathrm{R} - \langle \function{\sigma_{31'}}{t} \rangle_\mathrm{R}\right]  \right\} \function{\delta}{t-t'}\,.
    \end{align}
    \label{eq:cnumberdiffpop}
  \end{subequations}
\end{widetext}
\endgroup
{As an example, we provide a detailed derivation of the diffusion term from Eq.\eqref{eq:cnumberdiffpop}(a) in Appendix~\ref{sec:appendix-diff-coeff}.}
Additionally, we obtain diffusion coefficients absent in the quantum Langevin theory, e.g.
\begin{subequations}
  \begin{align}
    \langle \function{F_{23}}{t} \function{F_{23}}{t'}\rangle_\mathrm{R} = {}   & 2 \imu g \langle \function{a}{t} \function{\sigma_{23}}{t} \rangle_\mathrm{R}  \function{\delta}{t-t'}\,,
    \\
    \langle \function{F_{31'}}{t} \function{F_{31'}}{t'}\rangle_\mathrm{R} = {} & -2 \imu \Omega_{1'3} \langle \function{\sigma_{31'}}{t} \rangle_\mathrm{R}  \function{\delta}{t-t'}\,.
  \end{align}
\end{subequations}
The complete diffusion matrix $\function{\mat{D}}{\vec{A},t}$ including all relevant cross-correlation terms of the three-level QCL system with the c-number vector $\vec{A}=\left[a^*, a, \sigma_{23}^*, \sigma_{31'}^*, \sigma_{21'}^*, \sigma_{33}, \sigma_{22}, \sigma_{1'1'}, \sigma_{21'}, \sigma_{31'}, \sigma_{23}\right]^\transp$ is illustrated in Appendix~\ref{sec:appendix-diff-matrix}.

{In literature,\cite{drummond_hillery_2014,mandt2015stochastic} it has been shown that a set of Ito stochastic differential equations (SDEs) can be derived for the given c-number vector and can serve as an efficient basis for numerical simulations. The equivalent Ito SDEs to the Langevin theory are given by}
\begin{equation}
  \partial_t \function{\vec{A}}{t} = \function{\mat{M}}{t} + \function{\mat{F}}{t}
  = \function{\mat{M}}{t} + \function{\mat{B}}{\vec{A},t} \cdot
  \function{\vec{\xi}}{t}\,,
\end{equation}
where $\function{\vec{\xi}}{t}$ is a vector with real, independent Gaussian random numbers. {Here, a semi-definite and symmetric diffusion matrix $\function{\mat{D}}{\vec{A},t}$ is required, which then can be factorized into the form\cite{drummond_hillery_2014,drummond1980generalised,drummond1991quantum}}
\begin{equation}
  \function{\mat{D}}{\vec{A},t} = \function{\mat{B}}{\vec{A},t}\function{\mat{B}^\transp}{\vec{A},t}\,,
\end{equation}
where the derived noise matrix $\function{\mat{B}}{\vec{A},t}$ is not necessarily symmetric.

To calculate the full noise matrix $\function{\mat{B}}{\vec{A},t}$ for the
three-level QCL system, we can divide the diffusion matrix
$\function{\mat{D}}{\vec{A},t}$ into four different submatrices, where a
correlation between the corresponding terms is identified. The given
subvector $\vec{A}_\nu$ as well as the submatrices
$\function{\mat{B}_\nu}{\vec{A}_\nu,t}$ and
$\function{\mat{D}_\nu}{\vec{A}_\nu,t}$ are illustrated in
Table\thinspace\ref{tab:noise-matrix-terms}. Here, we include correlations
between three states by taking into account a tunneling transition followed
by an optical transition. This leads to a substantial extension of the
initially derived quantum theory of propagation of nonclassical radiation in
a two-level system\cite{drummond1991quantum} and is of essential importance
for the description of quantum fluctuations in THz QCL systems, where
electron transport across thick barriers is mediated by tunneling between
closely aligned energy levels. A detailed symbolic derivation of the noise
submatrices $\function{\mat{B}_\nu}{\vec{A}_\nu,t}$ and the resulting noise
matrix $\function{\mat{B}}{\vec{A},t}$ for the three-level QCL system can be
found in the GitHub project mbsolve.\cite{popp2023mbsolve}

\begin{table*}[btp]
  \caption{Division of the diffusion matrix into submatrices $\function{\mat{D}_\nu}{\vec{A}_\nu,t}$ and the corresponding c-number subvectors $\vec{A}_\nu$ and noise submatrices $\function{\mat{B}_\nu}{\vec{A}_\nu,t}$. In order to preserve the physical properties of the quantum system description, we have to interpret noise matrices  $\function{\mat{B}_\nu}{\vec{A}_\nu,t}$ differently for occupation and coherence terms. The differing matrix expressions for the coherence terms are highlighted here in red.}
  \centering
  \begin{ruledtabular}
    \begin{tabular}{ccc}
      subvector $\vec{A}_\nu$                                                                                                                                                                                                                                                                                                                                                                                                                                                                                                                                                                                              & submatrix $\function{\mat{B}_\nu}{\vec{A}_\nu,t}$                                                                                                  &
      submatrix $\function{\mat{D}_\nu}{\vec{A}_\nu,t}$                                                                                                                                                                                                                                                                                                                                                                                                                                                                                                                                                                                                                                                                                                                                                                                                                                                                                                                                                                              \\
      \hline \addlinespace[2ex]
      $\begin{bmatrix}\sigma_{23}^*                                                           \\ \sigma_{33}\\ \sigma_{22}\\ \sigma_{23}\end{bmatrix}$, $\begin{bmatrix}\sigma_{31'}^*                                                                                                                                                                                                                                                                        \\ \sigma_{33}\\ \sigma_{1'1'}\\ \sigma_{31'}\end{bmatrix}$, $\begin{bmatrix}\sigma_{21'}^* \\ \sigma_{22}\\ \sigma_{1'1'}\\ \sigma_{21'}\end{bmatrix}$,     & \multirow{2}{*}{$\begin{bmatrix} a   & -\imu a  \\ -b  & -\imu c \\ b   & \imu c   \\ a^* & \imu a^* \\ \end{bmatrix}$ }                           & \multirow{2}{*}{$ \begin{bmatrix} 0 & -ab -ac & ab + ac & 2\abs{a}^2 \\ -ab - ac   & b^2 -c^2 & - b^2 + c^2 & - a^*b  + a^*c \\ ab + ac & - b^2 + c^2 & b^2 - c^2 & a^*b -a^*c \\ 2\abs{a}^2 & - a^*b + a^*c & a^*b - a^*c & 0 \\ \end{bmatrix} $} \\ \addlinespace[2ex] $\begin{bmatrix}\sigma_{23}^* \\ \sigma_{22}\\ \sigma_{1'1'}\\ \sigma_{23}\end{bmatrix}$, $\begin{bmatrix}\sigma_{31'}^* \\ \sigma_{33}\\ \sigma_{22}\\ \sigma_{31'}\end{bmatrix}$, $\begin{bmatrix}\sigma_{21'}^* \\ \sigma_{33}\\ \sigma_{22}\\ \sigma_{21'}\end{bmatrix}$ && \\
      \addlinespace[2ex] \hline \addlinespace[2ex]

      $\begin{bmatrix}\sigma_{33}                                                             \\ \sigma_{22}\end{bmatrix}$, $\begin{bmatrix}\sigma_{33} \\ \sigma_{1'1'}\end{bmatrix}$, $\begin{bmatrix}\sigma_{22} \\ \sigma_{1'1'}\end{bmatrix}$                                                                                                                                                                                                                                                                                                                                                                         & \multirow{2}{*}{$ \begin{bmatrix}[*2l] a  \\ -a \\ \end{bmatrix} $ , \textcolor{TUMExtRed}{$ \begin{bmatrix}[*2l] a   \\ a^* \\ \end{bmatrix}  $}} & \multirow{2}{*}{$ \begin{bmatrix}[*2l] a^2  & -a^2 \\ -a^2 & a^2  \\ \end{bmatrix} $, \textcolor{TUMExtRed}{$ \begin{bmatrix}[*2l] a^2       & \abs{a}^2 \\ \abs{a}^2 & (a^*)^2   \\ \end{bmatrix} $}}                                             \\  \addlinespace[2ex]  \textcolor{TUMExtRed}{$\begin{bmatrix}\sigma_{23}^* \\ \sigma_{23}\end{bmatrix}$, $\begin{bmatrix}\sigma_{31'}^* \\ \sigma_{31'}\end{bmatrix}$} & & \\
      \addlinespace[2ex] \hline \addlinespace[2ex]
      $\begin{bmatrix}\sigma_{23}^*                                                           \\ \sigma_{31'}^*\\ \sigma_{31'}\\ \sigma_{23}\end{bmatrix}$, $\begin{bmatrix}\sigma_{23}^*                                                                                                                                                                                                                                                                         \\ \sigma_{21'}^*\\ \sigma_{21'}\\ \sigma_{23}\end{bmatrix}$, $\begin{bmatrix}\sigma_{23}^* \\ \sigma_{21'}\\ \sigma_{21'}^*\\ \sigma_{23}\end{bmatrix}$ & $ \begin{bmatrix} a & \imu a \\ b & -\imu b \\ c & \imu c \\ a^* & - \imu a^* \\ \end{bmatrix}$                                                    & $ \begin{bmatrix} 0 & 2ab & 0 & 2\abs{a}^2 \\ 2ab & 0 & 2bc & 0 \\ 0 & 2bc & 0 & 2a^*c \\ 2\abs{a}^2 & 0 & 2a^*c & 0 \\ \end{bmatrix} $                                                                                                            \\
      \addlinespace[2ex] \hline \addlinespace[2ex]
      $\begin{bmatrix}\sigma_{23}^*                                                           \\ \sigma_{23}\end{bmatrix}$, $\begin{bmatrix}\sigma_{31'}^* \\ \sigma_{31'}\end{bmatrix}$, $\begin{bmatrix}\sigma_{21'}^* \\ \sigma_{21'}\end{bmatrix}$                                                                                                                                                                                                                                                                                                                                                                     & $ \begin{bmatrix} a & \imu a  \\ a & -\imu a \\ \end{bmatrix}$                                                                                     & $ \begin{bmatrix} 0    & 2a^2 \\ 2a^2 & 0    \\ \end{bmatrix} $                                                                                                                                                                                    \\
      \addlinespace[2ex]
    \end{tabular}
  \end{ruledtabular}
  \label{tab:noise-matrix-terms}
\end{table*}

By calculating the operator expectation value in the Schrödinger picture; we
can demonstrate that the c-numbers representing the quantum system can be
replaced by the density matrix elements $\rho_{23},\, \rho_{31'},\,
  \rho_{21'},\, \rho_{33},\, \rho_{22},\, \rho_{1'1'},\, \rho_{1'2},\,
  \rho_{1'3},\, \rho_{32}$. The expectation value can be written as
\begin{equation}
  \begin{split}
    \langle \op{\sigma}_{ij} \rangle {}& = \traceu\left\lbrace|i\rangle\langle j | \function{\op{\rho}}{t} \right\rbrace \\ {}& = \traceu\left\lbrace|i\rangle\langle j |  \sum_{j',i'} \function{\rho_{j'i'}}{t} |j'\rangle\langle i' |  \right\rbrace = \function{\rho_{ji}}{t}\,.
  \end{split}
\end{equation}
Furthermore, we can write the interaction Hamiltonian $\op{H}_\mathrm{I}$ of the quantum system and the optical field as
\begin{equation}
  \op{H}_\mathrm{I} = -\op{\vec{\mu}}_z \op{\vec{E}}_z = -\vec{\mu}_{z,23} \op{\vec{E}}_z \left(\op{\sigma}_{32} + \op{\sigma}_{23}\right)\,,
\end{equation}
where the electrical field operator $\op{\vec{E}}_z$ is defined as
\begin{equation}
  \op{\vec{E}}_z = \sqrt{\frac{\omega_0}{2\hbar\epsilon_r\epsilon_0 V_\mathrm{p}}} \left(\op{a}^\dagger + \op{a}\right) \vec{e}_z \,.
\end{equation}
Here, $\vec{e}_z $ denotes the unit vector in $z$-direction. For devices in which the intraband transitions between quantized states occur within the conduction band, e.g. the QCL quantum well heterostructure,  only the dipole matrix element $\vec{\mu}_z$ for the polarization in growth direction $z$ is nonzero and relevant.

\subsection{Generalized Maxwell-density matrix Langevin equations in 1D}
In the following, we derive the generalized Maxwell-density matrix Langevin
equations with additional microscopic fluctuation terms and characterize the
influence of spontaneous emission noise on the optical field evolution. For
the description of the coherent carrier dynamics and the incoherent
relaxation processes, as well as the interaction with the classical optical
field, the generalized full-wave Maxwell-density matrix equations constitute
a compact semiclassical model. By combining it with the Langevin approach the
microscopic noise characteristics can be fully taken into account. Here, the
carrier dynamics in a SCL system are described in the density matrix
formulation using the Lindblad equation
\begin{equation}
  \partial_t{\op{\rho}}=-\imu\hbar^{-1}\commutator{\op{H}_0+\Delta
    \op{V}_\mathrm{tb} +
    \op{H}_\mathrm{I}}{\op{\rho}}+\function{\mathcal{D}}{\op{\rho}}+\function{\mathcal{F}}{\op{\rho}}\,,
  \label{eq:Lindblad}
\end{equation}
which is coupled to Maxwell's equations in one dimension
\begin{subequations}
  \label{eq:Maxwell}
  \begin{align}
    \partial_t{E_z} ={} & \varepsilon^{-1}\left(-\sigma E_z - \partial_t{P_{z,\mathrm{class}}} -\Gamma\partial_t{P_{z,\mathrm{qm}}}+\partial_x{H_y}\right)\,, \\
    \partial_t{H_y} ={} & \mu^{-1}\partial_x{E_z}\,,
  \end{align}
\end{subequations}
where $\function{\mathcal{D}}{\op{\rho}}$ is the dissipation superoperator,
$\function{\mathcal{F}}{\op{\rho}}$ is an additional Langevin fluctuation
superoperator and the other operators have their usual meanings. The
permittivity is given by the product $\varepsilon = \varepsilon_0
  \varepsilon_\mathrm{r}$, $\sigma$ is the material conductivity, $\mu$ is the
permeability, and the confinement factor $\Gamma \in [0,1]$ gives the spatial
overlap of the transverse optical field mode with the quantum system. As we
focus in this work on optoelectronic devices with invariant transverse field
distribution, the reduction to a one-dimensional model for the optical
propagation in the waveguide is justified.\cite{jirauschek2019optoelectronic}
The Lindblad equation is the general form of a time-local and Markovian
linear master equation for a quantum system, described by its completely
positive trace-preserving density matrix, interacting with an environment.
Obviously, the conventional Bloch equations, corresponding to a two-level
system, describe the interaction of the laser levels with the optical field
$E_z$ and constitute a special case of the Lindblad equation given in
Eq.~\eqref{eq:Lindblad}. The interaction with the environment is here modeled
by scattering and dephasing rates, $r_{ij}$ and $\gamma_{ij}$. Further levels
can be considered in Eq.~\eqref{eq:Lindblad}, and additional effects such as
tunneling are included in the Hamiltonian. Moreover, quantum fluctuations are
considered in the model given by Eq.~\eqref{eq:Lindblad} by adding a suitable
Langevin fluctuation superoperator $\mathcal{F}$. Maxwell's equations capture
the optical propagation through the waveguide resonator, where the coupling
with the quantum system is described by the macroscopic polarization $P_{z,
  \mathrm{qm}}$ arising from the contributions of the dipole matrix elements.
The expectation value of the dipole moment operator $\op{\mu}_z$ is
calculated by averaging over a large ensemble of quantum systems within an
adequate volume $V_p$ around the position $z$, and we can write for the
macroscopic polarization
\begin{equation}
  \label{eq:pol_qm}
  \begin{split}
    P_{z,\mathrm{qm}} {}& =  n_\mathrm{3D}\traceu\left\lbrace\op{\mu}_z {\op{\rho}}\right\rbrace = n_\mathrm{3D} \left( \mu_{z,23}\rho_{32} + \mu_{z,32}\rho_{23}\right) \\ & =n_\mathrm{3D} \mu_{z,23} \left( \rho_{32} + \rho_{23}\right)\,,
  \end{split}
\end{equation}
where $n_\mathrm{3D}$ is the carrier number density. The two classical
contributions, $P_{z, \mathrm{class}}= \epsilon_0 \chi E_z$ and $\sigma E_z$,
account for the polarization caused by bulk and waveguide dispersion as well
as the material losses.\cite{seitner2021group}

Finally, the update equations of the density matrix elements for the QCL
laser system depicted in Fig.~\ref{fig:wf_two_well} can be written as
\begingroup
\allowdisplaybreaks
\begin{subequations}
  \label{eq:density}
  \begin{align}
    \label{subeq:dens32}
    \partial_t \function{\rho_{32}}{t} ={}   & -\frac{\imu}{\hbar}\Delta_{32}
    \function{\rho_{32}}{t} -\gamma_{23}\function{\rho_{32}}{t} + \imu
    \Omega_{1'3} \function{\rho_{1'2}}{t} \nonumber                             \\  & + \frac{\imu}{\hbar}
    \mu_{z,23} E_z \left[\function{\rho_{33}}{t} - \function{\rho_{22}}{t}\right]
    + \function{F_{23}}{t} \,,                                                  \\
    \partial_t \function{\rho_{1'3}}{t} ={}  & -\frac{\imu}{\hbar}\Delta_{1'3}
    \function{\rho_{1'3}}{t} -\gamma_{1'3} \function{\rho_{1'3}}{t} + \imu
    \Omega_{1'3} \left[\function{\rho_{33}}{t} \right. \nonumber                \\  & \left.
         -\function{\rho_{1'1'}}{t} \right] +\frac{\imu}{\hbar} \mu_{z,23} E_z
    \function{\rho_{1'2}}{t} + \function{F_{31'}}{t} \,,                        \\
    \partial_t \function{\rho_{1'2}}{t} ={}  & -\frac{\imu}{\hbar}\Delta_{1'2}
    \function{\rho_{1'2}}{t} -\gamma_{1'2} \function{\rho_{1'2}}{t} + \imu
    \Omega_{1'3} \function{\rho _{32}}{t} \nonumber                             \\  & +\frac{\imu}{\hbar}
    \mu_{z,23} E_z \function{\rho_{1'3}}{t} + \function{F_{21'}}{t} \,,         \\
    \partial_t \function{\rho_{33}}{t} ={}   & - \frac{1}{\tau_3}
    \function{\rho_{33}}{t} + r_{32}\function{\rho_{22}}{t} +
    r_{31'}\function{\rho_{1'1'}}{t} \nonumber                                  \\  & - 2 \hbar^{-1} \mu_{z,23} E_z
    \Im{\function{\rho_{32}}{t}} \nonumber                                      \\  & - 2 \Omega_{1'3}
    \Im{\function{\rho_{1'3}}{t}} + \function{F_{33}}{t}\,,                     \\
    \partial_t \function{\rho_{22}}{t} ={}   & r_{23}\function{\rho_{33}}{t} -
    \frac{1}{\tau_2} \function{\rho_{22}}{t} + r_{21'}\function{\rho_{1'1'}}{t}
    \nonumber                                                                   \\  & + 2 \hbar^{-1} \mu_{z,23} E_z \Im{\function{\rho_{32}}{t}} +
    \function{F_{22}}{t}\,,                                                     \\
    \partial_t \function{\rho_{1'1'}}{t} ={} & r_{1'3}\function{\rho_{33}}{t} +
    r_{1'2}\function{\rho_{22}}{t} - \frac{1}{\tau_{1'}}
    \function{\rho_{1'1'}}{t} \nonumber                                         \\  & + 2 \Omega_{1'3}
    \Im{\function{\rho_{1'3}}{t}} + \function{F_{1'1'}}{t}\,.
  \end{align}
\end{subequations}
\endgroup
Via the macroscopic polarization $P_{z,\mathrm{qm}}$, the quantum
fluctuations added to the coherence term of Eq.~\eqref{subeq:dens32} have an
influence on the evolution of the classical optical field. The
quantum-mechanical fluctuations are based on the symbolic evaluations derived
in Sec.\thinspace\ref{subsec:c-numberLangevin} and can be found in
Appendix~\ref{sec:appendix-quant-fluc}. For the reduction to a two-level
system, we obtain similar noise terms as derived by Drummond and
Raymer.\cite{drummond1991quantum} However, unlike Drummond and Raymer we can
assure the preservation of the physical properties of the density matrix,
i.e., positive definiteness and unit trace. This is accomplished by a
suitable choice of the submatrices $\function{\mat{B}_\nu}{\vec{A}_\nu,t}$
depicted in Table \thinspace\ref{tab:noise-matrix-terms}.
\section{\label{sec:simulation}Simulation}
{The density matrix equations can be used to model the carrier dynamics in low-dimensional SCLs.\cite{kuhn1998} In QCLs, for example, optical intersubband transitions occur between quantized energy levels in the conduction band of a multiple quantum well heterostructure. The quantized energy state with index $i$ refers here to a subband and is characterized by a wavefunction $\psi_i$. Due to the 1D-confinement in growth direction, the free electron motion is restricted to the in-plane directions and is in carrier transport simulations characterized by an in-plane wavevector $\vec{k}=\left[k_x, k_y\right]^\transp$. Various 1D and 3D density matrix models for QCL structures are available in literature.\cite{iotti2001nature,iotti2005microscopic,terazzi2010density,weber2009density,wang2007coherent,dupont2010simplified,kumar2009coherence,freeman2013laser,jirauschek2023theory,popp2023self} In THz QCLs with small energy level spacings, we can typically neglect nonparabolicity effects and obtain $\vec{k}$ independent optical transition frequencies $\omega_{ij}$. In addition, Pauli blocking and Hartree-Fock renormalization effects, which occur in interband SCLs, play a subordinate role in QCLs due to the typically low doping levels.\cite{waldmueller2006nonequilibrium,jirauschek2019optoelectronic} Since dynamic simulations of nonlinear optical effects, such as the generation of frequency combs, have to run over thousands of cavity roundtrips to achieve convergence, a complete 3D Maxwell-density matrix simulation is not feasible from a numerical point of view. If we take into account the $\vec{k}$ conservation of the dipole matrix elements, we can sum over $\vec{k}$ which yields density matrix equations of the form in Eq.~\eqref{eq:density}. Under the assumption of moderate temporal variations of intersubband electron distributions $\rho_{jj,\vec{k}}$, averaged scattering rates $r_{ij}$ can be calculated by\cite{jirauschek2017self,jirauschek2019optoelectronic}
  \begin{equation}
    r_{ij} =  \left.\sum\limits_{\vec{k}, \vec{k'}}  r_{ j\vec{k'}\rightarrow i\vec{k}} \rho_{jj,\vec{k'}}^0 \right/  \sum\limits_{\vec{k}}  \rho_{jj,\vec{k'}}^0 \,,
  \end{equation}
  where $ \rho_{jj,\vec{k'}}^0 $ is the steady-state electron distribution in the subband $j$.  The $\vec{k}$ dependent scattering rates $r_{ j\vec{k'}\rightarrow i\vec{k}}$ can be extracted, e.g., from self-consistent DM-EMC simulations where they are calculated using Fermi's golden rule.\cite{jirauschek2014modeling} In QCLs, the electron distributions within one subband can often be well described by Fermi-Dirac or Maxwell-Boltzmann distributions. Here, the characteristic subband electron temperatures can significantly exceed the lattice temperatures.\cite{jirauschek2014modeling,vitiello2005measurement} The effective dephasing rates $\gamma_{ij}$ are typically obtained by averaging over the inversion between the subbands as \cite{nelander2009temperature,freeman2016self}
  \begin{equation}
    \gamma_{ij} =  \left.\sum\limits_{\vec{k}}  \gamma_{ ij,\vec{k'}} | \rho_{ii,\vec{k}}^0  - \rho_{jj,\vec{k}}^0 |\right/ \sum\limits_{\vec{k}} | \rho_{ii,\vec{k}}^0  - \rho_{jj,\vec{k}}^0 | \,.
  \end{equation}
}
As mentioned above, the Maxwell-density matrix equations are commonly treated
in the so-called rotating-wave/slowly varying amplitude approximation to
reduce the numerical load, which is only valid for relatively narrowband (and
not too strong) optical fields.\cite{allen1987optical,jirauschek2019optoelectronic,
  slavcheva2002coupled} However, as discussed in the introduction, in the
framework of this paper a broadband frequency comb with many modes is highly
beneficial, and QCLs even offer the potential for generating spectra
extending over a full octave and beyond.\cite{rosch2015octave} None of the
available open-source platforms are suitable for our purposes, mostly because
they employ the rotating-wave approximation. Thus, we have developed our own
open-source project mbsolve, allowing for numerically extensive simulations
of multilevel systems based on the full-wave Maxwell-density matrix
equations.\cite{popp2023mbsolve, riesch2021mbsolve} In detail, the
development of the codebase has been based on various principles. Here, the
generalized Lindblad equation [Eq.~\eqref{eq:Lindblad}] instead of the usual,
quite restrictive two-level Bloch equation model is used. We have developed
numerical methods that preserve physical properties, such as the complete
positivity and trace preservation of the density
matrix.\cite{jirauschek2019optoelectronic, riesch2019analyzing} This is
especially important in the context of long-term simulations, as required for
frequency comb modeling. Furthermore, a computational speedup is obtained by
using parallelization techniques,\cite{riesch2018performance} which is also
especially important for long-term simulations of rather complex quantum
systems, e.g. intracavity THz frequency comb difference frequency generation
by mid-IR QCLs.\cite{popp2023self} Our scientific software package mbsolve is
developed following sustainable software engineering strategies and includes
all common and essential best software engineering
practices.\cite{riesch2018performance,riesch2020bertha} It is based on C++
for performance reasons and features an easy-to-use Python interface
facilitating the setup and active quantum system of the low-dimensional
optoelectronic structures. The central part of the software is the
mbsolve-lib base library, providing a framework for defining a simulation
setup and the infrastructure to add solver and writer components.
Importantly, mbsolve supports different numerical methods for solving the
Lindblad equation,\cite{riesch2019analyzing} as well as different
parallelization techniques, e.g. OpenMP for shared memory systems. For a
detailed package description, the reader is referred to
Ref.~\onlinecite{riesch2021mbsolve}.

As pointed out above, we extended our code base considering vacuum
fluctuations due to spontaneous emission and fluctuations associated with the
electronic transport.\cite{andreasen2009finite,pusch2012coherent,
  slavcheva2004fdtd} The solver class \textbf{solver\_cpu\_fdtd} has the method
run, which executes the simulation loop by updating the magnetic field $H_y$,
electric field $E_z$, density matrix $\rho$ and polarization
$P_{z,\mathrm{qm}}$ for all spatial and temporal grid points in the Yee grid.
This update procedure is explained in detail in
Ref.~\onlinecite{riesch2021mbsolve}. We add a new density matrix algorithm
class, which is based on the Strang operator splitting
method,\cite{riesch2019analyzing, strang1968on} to account for fluctuations
accompanying the electronic transport and vacuum fluctuations. The density
algorithm class \textbf{algo\_lindblad\_reg\_cayley\_qnoise} contains the
method \textbf{propagate\_fluctuation}, which calculates the fluctuations for
an update step by adding the product of the fluctuation superoperator and the
time interval $\Delta t \cdot \mathcal{\function{F}{\rho}}$ to the updated
density matrix $\rho^{n+1/2}$. Here, we investigate active SCL gain media in
lasing operation above threshold. {In the simulations, we have to find a
  balance between numerical efficiency and modeling accuracy.} Most of the
fluctuation terms given in Sec.~\ref{sec:model} arise from the operator
ordering when reducing the operator equations to c-number Langevin equations.
As it was proven in
literature,\cite{Cerjan2015quantitative,andreasen2009finite} these terms are
negligible in the lasing regime above threshold with strong optical fields in
the laser cavity. The fluctuation terms for a $N$-level system featuring
diagonal elements $F_{ii}$ and off-diagonal elements $F_{ij} =
  F_{ji}^\dagger$ can thus be significantly reduced for the numerical treatment
and are described by \begingroup \allowdisplaybreaks
\begin{subequations}
  \begin{align}
    F_{ii}^j ={} & - F_{jj}^i = \xi_{1,ij} \sqrt{\frac{r_{ji} \rho_{ii} +  r_{ij} \rho_{jj}}{N_\mathrm{cell}}} \,,                                          \\
    F_{ii} ={}   & \sum_{j \neq i} F_{ii}^j                                                                      \,,                                        \\
    F_{ij} ={}   & \left(\xi_{2,ij} + \imu \xi_{3,ij}\right)\, \nonumber                                                                                    \\
                 & \times \sqrt{\frac{-\tau_j^{-1} \rho_{jj} + \sum_{n\neq j} r_{jn} \rho_{nn} + 2 \gamma_{ij} \rho_{jj}}{2 N_\mathrm{cell}}} \nonumber \,, \\  \quad\text{for } i >j,
  \end{align}
\end{subequations}
\endgroup
where $N_\mathrm{cell}$ is the number of carriers in one grid cell. The
$\xi_{2, ij}, \xi_{2,ij}$ and $\xi_{3,ij}$ are real Gaussian random numbers
and fulfill the correlation function
\begin{equation}
  \langle \xi_{k,ij}(t) | \xi_{l,mn}(t')\rangle = \delta_{kl}\delta_{im} \delta_{jn} \delta \left(t -t'\right) \,.
\end{equation}
{For future applications, in which a more detailed fluctuation treatment would be beneficial, it might be necessary to extend our numerical model by additional noise terms derived in the previous section.}
Furthermore, we have implemented a class \textbf{ic\_density\_random\_2lvl},
which represents random initial conditions for the common Maxwell-Bloch
two-level system. As the dipole moment operators $\sigma_{12},\, \sigma_{21}$
and the atomic operators $\sigma_{11},\, \sigma_{22}$ do not commute, we have
to take into account a non-vanishing initial stochastic value for the
polarization term following the uncertainty
principle.\cite{haake1979fluctuations,andreasen2009finite} The tipping angle
$\theta$ is obtained by drawing a random number from a Gaussian distribution
with a standard deviation $\sigma = 2N^{-1/2}_\mathrm{cell}$ and the angle
$\phi$ in xy-plane is obtained by drawing a random number from a uniform
distribution.\cite{pusch2012coherent}

\begin{figure}[ptb]
  \centering
  \includegraphics{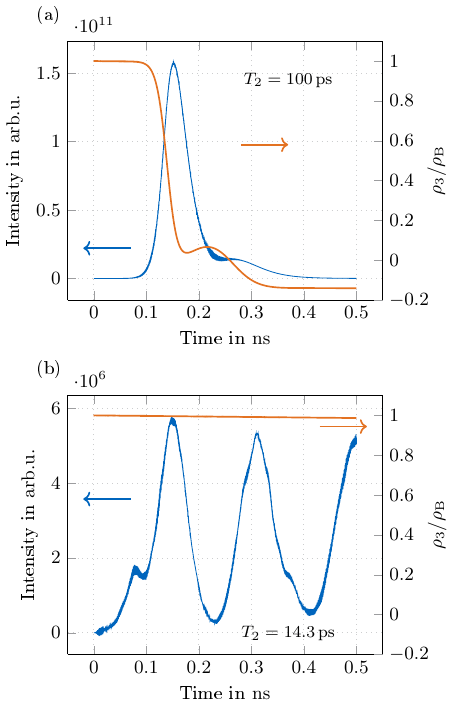}
  \caption{Simulation results for a superfluorescence test setup\cite{andreasen2009finite} in an initially inverted two-level system using the Maxwell-density matrix Langevin equations. (a) Cooperative emission characteristic of superfluorescence for the dephasing time $T_2 = \SI{100}{\pico\second}$. (b) Amplified spontaneous emission pulse for the dephasing time $T_2 = \SI{14.3}{\pico\second}$.}
  \label{fig:SF-pulse}
\end{figure}

\subsection{Superfluorescence and amplified spontaneous emission}
The system was tested with a superfluorescence (SF) setup in a two-level
configuration.\cite{andreasen2009finite,maki1989influence} This setup
describes the spontaneous build-up of a macroscopic coherent dipole moment in
an initially inverted system, resulting in a collective emission of a
superfluorescent pulse. This behavior can be reproduced numerically within
our mbsolve framework by simulating an ensemble of excited ions and using a
dephasing time $T_2 = \SI{100}{\pico\second}$. {All other parameters
  required for the simulation are taken from
  Ref.~\onlinecite{andreasen2009finite}. There, an excited state's lifetime
  $T_1=\SI{76}{\nano\second}$, carrier number density $n_\mathrm{3D}=
    \SI[per-mode=power]{8.53e19}{\per\cubic\meter}$, carrier number per cell
  $N_\mathrm{cell}= \SI{3e4}{}$, transition frequency $f =
    \SI{477}{\tera\hertz}$, dipole length $d=\SI{6.875e-2}{\nano\meter}$ and
  equilibrium inversion $w_0 = -1$ are specified. We further investigate a
  device with length $L = \SI{7}{\milli\meter}$ using a grid discretization
  $\Delta x = \SI{70}{\nano\meter}$.}

The simulated SF pulse is illustrated in Fig.~\ref{fig:SF-pulse}(a), and
compares well with previous numerical and experimental
findings.\cite{andreasen2009finite,malcuit1987transition,maki1989influence}
By increasing the collisional dephasing rate within the system, the SF pulse
is significantly disturbed and gets broadened until the spontaneous build-up
of the coherent dipole moment is prevented. For a dephasing time $T_2 =
  \SI{14.3}{\pico\second}$ below the critical point, the SF pulse is replaced
by amplified spontaneous emission (ASE). The increased noise amplitude
accompanying the smaller dephasing time is crucial for the modeling of ASE,
which cannot be reproduced otherwise. The ASE simulation results are
presented in Fig.~\ref{fig:SF-pulse}(b).

{Furthermore, the degree of decoherence is studied using the quantity
$\rho_3/\rho_\mathrm{B}$, where $\rho_\mathrm{B} = \sqrt{\rho_1^2+\rho_2^2+\rho_3^2}$ is the length of the Bloch vector. When the dephasing time $T_2$ is high [Fig.~\ref{fig:SF-pulse}(a)], the population inversion
$\rho_3$ is quickly depleted through the spontaneous buildup of the
macroscopic dipole moment and the SF emission, which clearly surpasses the
decay of $\rho_1$ and $\rho_2$ and results in a rapid drop of
$\rho_3/\rho_\mathrm{B}$. In the second case [Fig.~\ref{fig:SF-pulse}(b)] we
have used a smaller dephasing time $T_2$, which prevents the macroscopic
dipole moment build-up and limits the radiative decay. This decoherence state
indicates a very slow decay of $\rho_3/\rho_\mathrm{B}$, which stays close to
one.}
\subsection{\label{sec:hc_noise} Noise characteristics in THz QCL harmonic frequency comb emission} Concerning
the experimental investigation of intensity correlations in
QCLs,\cite{gabbrielli2022intensity} we aim to characterize the noise
properties of a self-starting THz QCL HFC setup. Here, the THz QCL active
region is based on a homogeneous four-quantum well design with a diagonal
transition.\cite{forrer2021self} The charge carrier transport in the active
gain medium at a bias of \SI{50}{\kilo\volt\per\period} is analyzed using our
in-house Monaco framework, consisting of a Schrödinger-Poisson solver and a
density matrix-ensemble Monte Carlo modeling
tool.\cite{jirauschek2009accuracy,jirauschek2014modeling,
  jirauschek2017self,jirauschek2017density,popp2021bayesian} For an appropriate
description of the physical properties, we consider five wavefunctions in the
active quantum well heterostructure. Furthermore, one incoherent tunneling
transition from the injector state into the upper laser level and one optical
transition are specified for the quantum mechanical description of the QCL
system in the dynamical simulation. {The Python script
  \textbf{forrer\_2021\_50mVperperiod.py} with the simulation setup to start
  the mbsolve simulation is given in Listing~\ref{lst:forrer2021} of
  Appendix~\ref{sec:mboslve-python} and is furthermore included in the GitHub
  repository.\cite{popp2023mbsolve} Here, all input parameters for the full
  description of the quantum system are extracted from self-consistent DM-EMC
  simulations.}
\begin{figure}[htp]
  \includegraphics{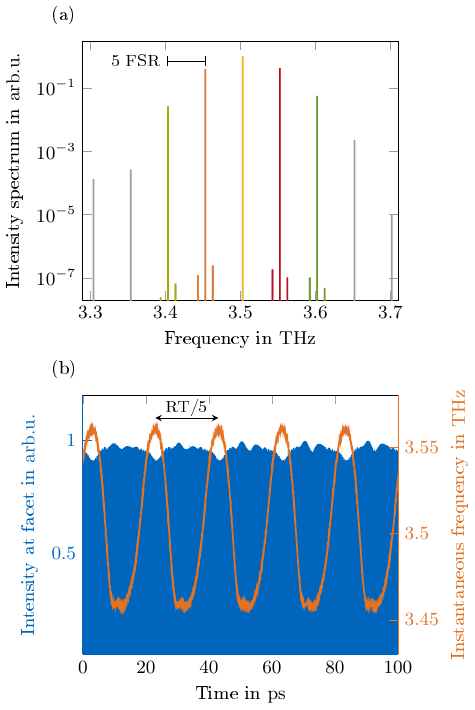}

  \caption{Maxwell-density matrix Langevin simulation results of HFC emission with a mode spacing of 5 FSR in a \SI{4}{\milli\meter} long THz QCL device with a metal-metal waveguide at \SI{80}{\kelvin} and for $V=\SI{50}{\milli\volt\per\period}$. (a) Intensity
    spectra of the optical radiation at the facet. (b) Simulated instantaneous intensity at the facet and calculated instantaneous frequency from the Hilbert transform of the simulated electric field over a single roundtrip time (RT). }
  \label{fig:mb-results}
\end{figure}

In the following, we present simulation results for a \SI{4}{\milli\meter}
long double-metal THz QCL with a free spectral range (FSR) of
\SI{9.94}{\giga\hertz}. The intensity spectrum of the THZ HFC at
\SI{3.5}{\tera\hertz} with a mode spacing of 5 FSR is illustrated in
Fig.~\ref{fig:mb-results}(a). The THz QCL emits a broadband HFC with a cavity
repetition rate of \SI{49.7}{\giga\hertz}. In Fig.~\ref{fig:mb-results}(b),
the temporal evolution of the intensity at the facet and the calculated
instantaneous frequency are depicted. We can identify a regular field
pattern, which shows a periodic repetition with five times the roundtrip
time. Here, only the three strongest modes are involved in the temporal
evolution of the instantaneous frequency, as their intensities are of similar
magnitude and contribute most to the overall comb emission power.
\begin{figure}
  \includegraphics{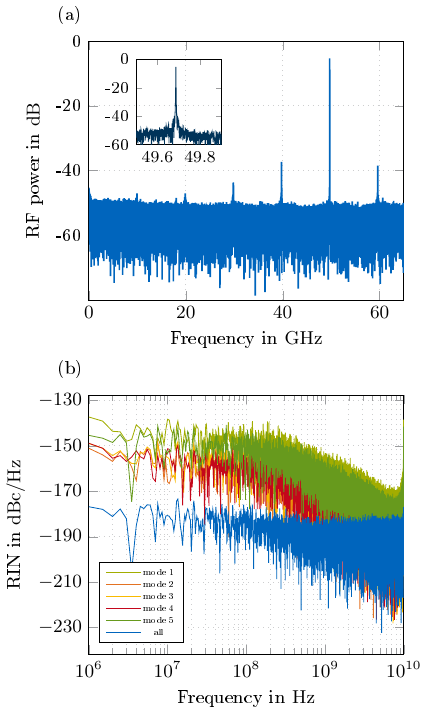}

  \caption{(a) Simulated RF spectrum of the THz QCL HFC setup with a clear beatnote signal at \SI{49.7}{\giga\hertz}. Inset, zoom on the harmonic beatnote, indicating a narrow linewidth below the numerical frequency resolution (\SI{500}{\kilo\hertz}). (b) Calculated RIN spectra associated with the total power $P_\mathrm{all}$ (blue) and the modal power $P_i$ of each of the five harmonic modes contributing most to the HFC emission. The colors of the individual RIN spectra correspond to those of the individual comb lines in Fig.~\ref{fig:mb-results}(a).}
  \label{fig:RF-RIN}
\end{figure}

To specify the degree of coherence of the obtained HFC and for comparison
with the experimental findings, we investigate the radio frequency (RF)
spectrum using an observation time window of $\SI{2}{\upmu\second}$. The
obtained simulation results are shown in Fig.~\ref{fig:RF-RIN}(a), and the
clear appearance of the harmonic beatnote proves the purity of the harmonic
state. The linewidth is substantially below the numerical frequency
resolution of \SI{500}{\kilo\hertz}, which is confirmed by the zoom on the
extremely narrow harmonic beatnote in the inset of Fig.~\ref{fig:RF-RIN}(a).
In addition, we can identify sub-beatnotes, which arise due to the beating of
the center mode with the sub-comb lines. These sub-comb lines are generated
by FWM processes, where the strong harmonic sidemodes act as pump modes and
generate weak sidebands with a frequency spacing of 1 FSR from the
corresponding pump modes. As can be seen in Fig.~\ref{fig:mb-results}(a), the
intensities of the sub-modes are at least $\sim 5$ orders of magnitude
smaller than those of the pump modes.

To further analyze the noise characteristics of the THz QCL HFC setup, we
calculate the relative intensity noise (RIN) for the total output power and
for the power of the five harmonic comb lines contributing most to the HFC
emission. Here, the RIN spectrum can be calculated by
\begin{equation}
  \begin{split}
    \function{\mathrm{RIN}_i}{f} = &\lim_{T\to\infty }\frac{1}{T} \frac{|\int^T_0 \left[ \function{P_i}{t} -  < \function{P_i}{t}> \right] \e^{-\imu 2\pi f t} \diff t |^2}{ < \function{P_i}{t}>^2}\,,
  \end{split}
\end{equation}
where $T$ denotes the simulation time, and $P_i$ is either the power of a
specific mode $i$ or the total power $P_\mathrm{all}$. By numerically
filtering the electric field at the facet $\function{E_\mathrm{facet}}{t}$
using a filter with a \SI{3}{\decibel} bandwidth of \SI{20}{\giga\hertz}, we
can extract the temporal electric field components $\function{E_i}{t}$ of the
individual modes. The RIN results are depicted in Fig.~\ref{fig:RF-RIN}(b)
for the total power $P_\mathrm{all}$ and the power of the five central
harmonic modes $P_i$ with indices $i = 1\ldots5$. The total power RIN is
around \SI{-180}{\dBc\per\hertz}, while for the three central harmonic modes
having a similar power a RIN around \SI{-155}{\dBc\per\hertz} is calculated.
For the remaining two weaker modes 1 and 5 a higher RIN is obtained. This is
in very good agreement with the experimental findings of a three-mode mid-IR
HFC QCL setup.\cite{gabbrielli2022intensity} For increasing power, the RIN of
the sidemodes decreases to that of the central mode, while sidebands closer
to threshold exhibit a noisier behavior. Furthermore, we identify an
overlapping RIN for sidemodes featuring a comparable power level, which
indicates a comparable noise level. A similar result could be retrieved from
the mid-IR HFC RIN measurements.\cite{gabbrielli2022intensity}
\section{\label{sec:conclusion}Conclusion} In this paper, we have theoretically derived a c-number Langevin
approach for a three-level quantum system from the non-classical operator
description within the Heisenberg-Langevin equations. Our approach is an
extension of the well-known two-level quantum theory by Drummond and
Raymer,\cite{drummond1991quantum} where we additionally take into account
incoherent tunneling injection into the upper laser level. Within the
generalized Maxwell-density matrix Langevin equations we can ensure the
preservation of the physical properties of the density matrix, i.e., positive
definiteness and unit trace. Furthermore, by including the derived noise
terms into our open-source simulation tool mbsolve, we can model the
fluctuations accompanying electronic transport and spontaneous emission in
the dynamical simulations of light-matter interaction in multilevel quantum
optoelectronic systems such as QCLs and QD lasers. The simulation approach is
tested using a superfluorescence setup, where we prove the validity of our
implementation by obtaining an excellent match with previous experimental and
theoretical results.\cite{andreasen2009finite,
  pusch2012coherent,maki1989influence,malcuit1987transition} Additionally, we
have characterized the noise properties of a coherent THz QCL HFC setup and
obtained a good match with experimental
findings.\cite{forrer2021self,gabbrielli2022intensity} Our modeling approach
based on the generalized Maxwell-density matrix Langevin equations shows
great potential for the theoretical investigation of intermodal intensity
correlations in photonic devices and the development of low-noise integrated
light emitters, also with regard to the generation of non-classical light.
\begin{acknowledgments}
  The authors acknowledge financial support by the European Union's QuantERA II [G.A. n. 101017733] - QATACOMB Project "Quantum correlations in terahertz QCL combs" (Funding organization: DFG - Germany [Project n. 491801597]), by the European Union's Research and Innovation Programmes Horizon 2020 and Horizon Europe with the Qombs Project [G.A. n. 820419] "Quantum simulation and entanglement engineering in quantum cascade laser frequency combs", {by Deutsche Forschungsgemeinschaft (DFG) under the DFG DACH project [Project No. 471090402], by the FWF Project I 5682-N "Cavity-assisted non-classical light generation" and by the ESA (Discovery EISI) Project 4000142337: "Simulation toolbox for unconditionally secure on-chip satellite quantum communication networks operating in the telecom wavelength range".}
\end{acknowledgments}

\section*{Data availability}
The code and all scripts for reproducing the data presented in this paper
are available on GitHub.

\appendix
\section{Calculation of the diffusion coefficient within the quantum and the c-number Langevin treatment}
\label{sec:appendix-diff-coeff}
\begingroup
\allowdisplaybreaks
By taking into account the orthogonality of the levels $\langle i | j\rangle
  = \delta_{ij}$, we obtain

\begin{equation}
  \op{\sigma}_{23}^\dagger\op{\sigma}_{23} = \left( |3\rangle \langle 2 | 2 \rangle \langle 3 | \right) = \left(|3\rangle \langle 3|\right)  = \op{\sigma}_{33}\,.
  \label{eq:operator_ortho}
\end{equation}
The diffusion coefficient in Eq.~\eqref{eq:d3223} is calculated by using the \textit{generalized Einstein relation} from Eq.~\eqref{eq:einstein}. The detailed calculation is given by
\begin{widetext}
  \begin{equation}
    \begin{split}
      2\langle \function{\op{D}_{3223}}{t} \rangle_\mathrm{R} ={} & \partial_t \langle \function{\op{\sigma}_{33}}{t} \rangle_\mathrm{R} - \langle
      \function{\op{M}_{23}^\dagger}{t}
      \function{\op{\sigma}_{23}}{t}\rangle_\mathrm{R} - \langle
      \function{\op{\sigma}_{23}^\dagger}{t} \function{\op{M}_{23}}{t}
      \rangle_\mathrm{R} \\
      ={} & \bigg \langle - \frac{1}{\tau_3} \function{\op{\sigma}_{33}}{t} + r_{32}\function{\op{\sigma}_{22}}{t} + r_{31'}\function{\op{\sigma}_{1'1'}}{t} + \imu g\left[ \function{\op{a}^\dagger}{t}\function{\op{\sigma}_{23}}{t} - \function{\op{a}}{t}\function{\op{\sigma}_{23}^\dagger}{t}\right] - \imu \Omega_{1'3} \left[ \function{\op{\sigma}_{31'}^\dagger}{t} - \function{\op{\sigma}_{31'}}{t}\right] \bigg \rangle_\mathrm{R} \\
      & - \bigg \langle \bigg\{\frac{\imu}{\hbar}\Delta_{32} \function{\op{\sigma}_{23}^\dagger}{t} -\gamma_{23}\function{\op{\sigma}_{23}^\dagger}{t} - \imu g \left[\function{\op{\sigma}_{33}}{t} - \function{\op{\sigma}_{22}}{t}\right] \function{\op{a}^\dagger}{t} - \imu \Omega_{1'3} \function{\op{\sigma}_{1'2}}{t} \bigg\} \function{\op{\sigma}_{23}}{t} \bigg \rangle_\mathrm{R}  \\
      & - \bigg \langle \function{\op{\sigma}_{23}^\dagger}{t}\bigg\{-\frac{\imu}{\hbar}\Delta_{32} \function{\op{\sigma}_{23}}{t} -\gamma_{23}\function{\op{\sigma}_{23}}{t} + \imu g \left[\function{\op{\sigma}_{33}}{t} - \function{\op{\sigma}_{22}}{t}\right] \function{\op{a}}{t} + \imu \Omega_{1'3} \function{\op{\sigma}_{21'}}{t} \bigg\} \bigg \rangle_\mathrm{R}\\
      ={} & \left(2\gamma_{23} - \frac{1}{\tau_3}\right)\langle\function{\op{\sigma}_{33}}{t}\rangle_\mathrm{R} + r_{32} \langle\function{\op{\sigma}_{22}}{t}\rangle_\mathrm{R} + r_{31'} \langle\function{\op{\sigma}_{1'1'}}{t}\rangle_\mathrm{R} \,.
    \end{split}
  \end{equation}
\end{widetext}
{In addition, we prove the difference in diffusion coefficients, which arises through the transition from operator to c-number Langevin equations. Therefore, we calculate the diffusion coefficient $\function{D_{3333}}{t}$ based on the theoretical description presented in Sec.~\ref{subsec:c-numberLangevin}. By the use of Eqs.~\eqref{eq:einstein} and \eqref{eq:Heisenberg}(e), we obtain
\begin{widetext}
  \begin{equation}
    \begin{split}
      \partial_t \langle\function{\op{\sigma}_{33}}{t} \function{\op{\sigma}_{33}}{t}
      \rangle_\mathrm{R} ={} & - \frac{2}{\tau_3}
      \langle\function{\op{\sigma}_{33}}{t}\function{\op{\sigma}_{33}}{t}\rangle_\mathrm{R}
      + r_{32} \left[\rulred{\langle\function{\op{\sigma}_{22}}{t}
          \function{\op{\sigma}_{33}}{t} \rangle_\mathrm{R}} +
        \langle\function{\op{\sigma}_{33}}{t}\function{\op{\sigma}_{22}}{t}
        \rangle_\mathrm{R} \right] +
      r_{31'}\Bigl[\rulred{\langle\function{\op{\sigma}_{1'1'}}{t}
          \function{\op{\sigma}_{33}}{t} \rangle_\mathrm{R}} \\ &
        +\langle\function{\op{\sigma}_{33}}{t}\function{\op{\sigma}_{1'1'}}{t}
        \rangle_\mathrm{R} \Bigr] + \imu g \left[\rulred{\langle
          \function{\op{a}^\dagger}{t}\function{\op{\sigma}_{23}}{t}\function{\op{\sigma}_{33}}{t}
          \rangle_\mathrm{R}} - \langle
        \function{\op{a}}{t}\function{\op{\sigma}_{23}^\dagger}{t}
        \function{\op{\sigma}_{33}}{t}\rangle_\mathrm{R} +
        \langle\function{\op{a}^\dagger}{t}\function{\op{\sigma}_{33}}{t}\function{\op{\sigma}_{23}}{t}
        \rangle_\mathrm{R} \right. \\ & \left. - \rulred{\langle
          \function{\op{a}}{t}\function{\op{\sigma}_{33}}{t}
          \function{\op{\sigma}_{23}^\dagger}{t}\rangle_\mathrm{R}}\right] - \imu
      \Omega_{1'3} \Bigl[ \langle \function{\op{\sigma}_{31'}^\dagger}{t}
        \function{\op{\sigma}_{33}}{t} \rangle_\mathrm{R} - \rulred{\langle
          \function{\op{\sigma}_{31'}}{t} \function{\op{\sigma}_{33}}{t}
          \rangle_\mathrm{R} +
          \langle\function{\op{\sigma}_{33}}{t}\function{\op{\sigma}_{31'}^\dagger}{t}
          \rangle_\mathrm{R}} \\ & - \langle\function{\op{\sigma}_{33}}{t}
        \function{\op{\sigma}_{31'}}{t}\rangle_\mathrm{R}\Bigr] + 2\langle
      \function{\op{D}_{3333}}{t} \rangle_\mathrm{R}\,.
    \end{split}
    \label{eq:eqmotion3333}
  \end{equation}
\end{widetext}
Here, the terms which are underlined in red are not in the chosen order defined in Sec.~\ref{subsec:c-numberLangevin}. The commutation relations are used to bring these terms into chosen order, and by exploiting the level orthogonality similar to Eq.~\eqref{eq:operator_ortho} we derive
\begin{subequations}
  \begin{align}
    \commutator{\function{\op{\sigma}_{22}}{t}}{\function{\op{\sigma}_{33}}{t}} = {}          & \function{\op{\sigma}_{22}}{t}\function{\op{\sigma}_{33}}{t} - \function{\op{\sigma}_{33}}{t}\function{\op{\sigma}_{22}}{t} \nonumber \\  = & 0 \,, \\
    \commutator{\function{\op{\sigma}_{23}}{t}}{\function{\op{\sigma}_{33}}{t}} = {}          & \function{\op{\sigma}_{23}}{t}               \,,                                                                                      \\
    \commutator{\function{\op{\sigma}_{33}}{t}}{\function{\op{\sigma}_{23}^\dagger}{t}} = {}  & \function{\op{\sigma}_{23}^\dagger}{t}        \,,                                                                                     \\
    \commutator{\function{\op{\sigma}_{31'}}{t}}{\function{\op{\sigma}_{33}}{t}} = {}         & -\function{\op{\sigma}_{31'}}{t}         \,,                                                                                          \\
    \commutator{\function{\op{\sigma}_{33}}{t}}{\function{\op{\sigma}_{31'}^\dagger}{t}} = {} & -\function{\op{\sigma}_{31'}^\dagger}{t} \,.
  \end{align}
\end{subequations}
With this, we can restructure Eq.~(\ref{eq:eqmotion3333}) as follows:
\begin{widetext}
  \begin{equation}
    \begin{split}
      \partial_t \langle\function{\op{\sigma}_{33}}{t} \function{\op{\sigma}_{33}}{t}
      \rangle_\mathrm{R} ={} & - \frac{2}{\tau_3}
      \langle\function{\op{\sigma}_{33}}{t}\function{\op{\sigma}_{33}}{t}\rangle_\mathrm{R}
      + 2r_{32} \langle\function{\op{\sigma}_{33}}{t}
      \function{\op{\sigma}_{22}}{t} \rangle_\mathrm{R} +
      2r_{31'}\langle\function{\op{\sigma}_{33}}{t}
      \function{\op{\sigma}_{1'1'}}{t} \rangle_\mathrm{R} \\ & + 2\imu g \left[
        \langle\function{\op{a}^\dagger}{t}\function{\op{\sigma}_{33}}{t}\function{\op{\sigma}_{23}}{t}
        \rangle_\mathrm{R} - \langle
        \function{\op{a}}{t}\function{\op{\sigma}_{23}^\dagger}{t}
        \function{\op{\sigma}_{33}}{t}\rangle_\mathrm{R} \right] - 2\imu \Omega_{1'3}
      \left[ \langle \function{\op{\sigma}_{31'}^\dagger}{t}
        \function{\op{\sigma}_{33}}{t} \rangle_\mathrm{R} -
        \langle\function{\op{\sigma}_{33}}{t}
        \function{\op{\sigma}_{31'}}{t}\rangle_\mathrm{R}\right] \\ &+ 2\langle
      \function{\op{D}_{3333}}{t} \rangle_\mathrm{R} + \rulgreen{ \imu g
        \left[\langle \function{\op{a}^\dagger}{t}\function{\op{\sigma}_{23}}{t}
          \rangle_\mathrm{R} - \langle \function{\op{a}}{t}
          \function{\op{\sigma}_{23}^\dagger}{t}\rangle_\mathrm{R} \right] + \imu
        \Omega_{1'3}\left[\langle\function{\op{\sigma}_{31'}^\dagger}{t}\rangle_\mathrm{R}
          - \langle \function{\op{\sigma}_{31'}}{t}\rangle_\mathrm{R}\right]}\,.
    \end{split}
    \label{eq:eqmotion3333-2}
  \end{equation}
\end{widetext}
Here, the additional terms resulting from the operator ordering are underlined in green. With the use of Eqs.~\eqref{eq:c-numberLangevinEq} and \eqref{eq:c-numberLangevinEq2} we can derive the corresponding c-number equation
\begin{widetext}
  \begin{equation}
    \begin{split}
      \partial_t \langle\function{{\sigma}_{33}}{t} \function{{\sigma}_{33}}{t}
      \rangle_\mathrm{R} ={} & - \frac{2}{\tau_3}
      \langle\function{{\sigma}_{33}}{t}\function{{\sigma}_{33}}{t}\rangle_\mathrm{R}
      + 2r_{32} \langle\function{{\sigma}_{33}}{t} \function{{\sigma}_{22}}{t}
      \rangle_\mathrm{R} + 2r_{31'}\langle\function{{\sigma}_{33}}{t}
      \function{{\sigma}_{1'1'}}{t} \rangle_\mathrm{R} + 2\imu g \left[
        \langle\function{{a}^*}{t}\function{{\sigma}_{33}}{t}\function{{\sigma}_{23}}{t}
        \rangle_\mathrm{R} \right. \\ & \left. - \langle
        \function{{a}}{t}\function{{\sigma}_{23}^*}{t}
        \function{{\sigma}_{33}}{t}\rangle_\mathrm{R} \right] - 2\imu \Omega_{1'3}
      \left[ \langle \function{{\sigma}_{31'}^*}{t} \function{{\sigma}_{33}}{t}
        \rangle_\mathrm{R} - \langle\function{{\sigma}_{33}}{t}
        \function{{\sigma}_{31'}}{t}\rangle_\mathrm{R}\right] + {2\langle
        \function{{D}_{3333}}{t} \rangle_\mathrm{R}}\,.
    \end{split}
    \label{eq:eqcnumber3333}
  \end{equation}
\end{widetext}
If we now require the equivalence of the left-hand sides of Eqs.~\eqref{eq:eqmotion3333-2} and \eqref{eq:eqcnumber3333}, we end up with the diffusion coefficient from Eq.\eqref{eq:cnumberdiffpop}(a).}
\endgroup
\section{Complete diffusion matrix for a three-level THz QCL system}
\label{sec:appendix-diff-matrix}
The complete diffusion matrix for a three-level THz QCL system with
incoherent tunneling injection is derived within the framework of the c-number Langevin theory and is given by
\begingroup
\allowdisplaybreaks
\renewcommand*{\arraystretch}{1.5}
\begin{widetext}
  \begin{align}
    A = &
    \left[
      \begin{matrix}
        0                          & \sqrt{n_\mathrm{th}\kappa} & 0                                                                                                                          & 0                                                                                                                                                     & 0                                                                                                                                                 \\
        \sqrt{n_\mathrm{th}\kappa} & 0                          & 0                                                                                                                          & 0                                                                                                                                                     & 0                                                                                                                                                 \\
        0                          & 0                          & -2 \imu g a^* \sigma_{23}^*                                                                                                & \imu g a^* \sigma_{31'}^*                                                                                                                             & -\imu g a^* \sigma_{21'}^*                                                                                                                        \\
        0                          & 0                          & \imu g a^* \sigma_{31'}^*                                                                                                  & 2 \imu \Omega_{1'3} \sigma_{31'}^*                                                                                                                    & 0                                                                                                                                                 \\
        0                          & 0                          & - \imu g a^* \sigma_{21'}^*                                                                                                & 0                                                                                                                                                     & 0                                                                                                                                                 \\
        0                          & 0                          & -r_{32} \sigma_{23}^*                                                                                                      & -\imu g a \sigma_{21'}^*  + \left(r_{23}  +r_{1'3} \right) \sigma_{31'}^*                                                                             & -r_{32} \sigma_{21'}^*                                                                                                                            \\
        0                          & 0                          & \left( r_{32} + r_{1'2} \right) \sigma_{23}^*                                                                              & \imu g a \sigma_{21'}^*  - r_{23} \sigma_{31'}^*                                                                                                      & \left(r_{32} + r_{1'2}\right) \sigma_{21'}^*                                                                                                      \\
        0                          & 0                          & - r_{1'2} \sigma_{23}^*                                                                                                    & -r_{1'3} \sigma_{31'}                                                                                                                                 & -r_{1'2} \sigma_{21'}^*                                                                                                                           \\
        0                          & 0                          & \left( \gamma_{23} + \gamma_{1'2} -\gamma_{1'3}\right) \sigma_{31'}                                                        & 0                                                                                                                                                     & \begin{smallmatrix} \left(2 \gamma_{1'2} - r_{21'} -r_{31'}\right) \sigma_{1'1'} \\ + r_{1'2} \sigma_{22} + r_{1'3} \sigma_{33} \end{smallmatrix} \\
        0                          & 0                          & 0                                                                                                                          & \begin{smallmatrix} \left(2\gamma_{1'3} - r_{21'} -r_{31'}\right) \sigma_{1'1'} + {} & \\ r_{1'2} \sigma_{22} + r_{1'3} \sigma_{33} \end{smallmatrix} & 0                                                                                                                                                 \\
        0                          & 0                          & \begin{smallmatrix} \left(2\gamma_{23} - r_{23} -r_{1'3}\right) \sigma_{33} {} & \\ + r_{32} \sigma_{22} \end{smallmatrix} & 0                                                                                                                                                     & \left(\gamma_{1'2} + \gamma_{23} - \gamma_{1'3} \right) \sigma_{31'}^*
      \end{matrix}
    \right.  \nonumber                                                                                                                                                                                                                                                                                                                                                                                                                                                                                                                                                                                                                                                                                                                                                                                                                                                                                         \\[1em]
        &
    \begin{matrix}
      0                                                                                                                                                                                                                                                     & 0                                                                                                                                                                                                                                                                                                                                                                                                                                              & 0                                                                                                                                                                                                 \\
      0                                                                                                                                                                                                                                                     & 0                                                                                                                                                                                                                                                                                                                                                                                                                                              & 0                                                                                                                                                                                                 \\
      - r_{32} \sigma_{23}^*                                                                                                                                                                                                                                & \left( r_{32} + r_{1'2} \right) \sigma_{23}^*                                                                                                                                                                                                                                                                                                                                                                                                  & -r_{1'2} \sigma_{23}^*                                                                                                                                                                            \\
      - \imu g a \sigma_{21'}^*  + \left(r_{23} +  r_{1'3} \right) \sigma_{31'}^*                                                                                                                                                                           & \imu g a \sigma_{21'}^*  - r_{23} \sigma_{31'}^*                                                                                                                                                                                                                                                                                                                                                                                               & -r_{1'3}\sigma_{31'}^*                                                                                                                                                                            \\
      -r_{32} \sigma_{21'}^*                                                                                                                                                                                                                                & \left( r_{32} + r_{1'2}\right) \sigma_{21'}^*                                                                                                                                                                                                                                                                                                                                                                                                  & - r_{1'2} \sigma_{21'}^*                                                                                                                                                                          \\
      \begin{smallmatrix} \left(r_{23} +r_{1'3}\right) \sigma_{33} + r_{32} \sigma_{22} + r_{3'1} \sigma_{1'1'} \\ + \imu g \left(a^*\sigma_{23} - a \sigma_{23}^*\right) + \imu \Omega_{1'3} \left( \sigma_{31'}^*  -\sigma_{31'}\right) \end{smallmatrix} & \begin{smallmatrix} -r_{32} \sigma_{22} - r_{23} \sigma_{33} \\+ \imu g \left( a\sigma_{23}^* - a^* \sigma_{23}\right)                           \end{smallmatrix}                                                                                                                                                                                                                                                                             & \begin{smallmatrix} \imu \Omega_{1'3} \left(\sigma_{31'}-\sigma_{31'}^*\right)\\ -r_{31'} \sigma_{1'1'} - r_{1'3} \sigma_{33}                          \end{smallmatrix}                          \\
      \begin{smallmatrix} -r_{32} \sigma_{22} - r_{23}\sigma_{33}\\ + \imu g \left( a\sigma_{23}^* - a^* \sigma_{23}\right) \end{smallmatrix}                                                                                                               & \begin{smallmatrix} r_{23} \sigma_{33} + \left( r_{32} + r_{1'2} \right) \sigma_{22} {}\\+  r_{21'} \sigma_{1'1'} +  \imu g \left( a^* \sigma_{23} - a \sigma_{23}^*\right)                                                                                                                                                                                                                                                  \end{smallmatrix} & -r_{21'} \sigma_{1'1'} - r_{1'2} \sigma_{22}                                                                                                                                                      \\
      \begin{smallmatrix} \imu \Omega_{1'3} \left(\sigma_{31'} - \sigma_{31'}^* \right) \\ -r_{1'3} \sigma_{33} -r_{31'} \sigma_{1'1'}  \end{smallmatrix}                                                                                                   & -r_{21'} \sigma_{1'1'} - r_{1'2} \sigma_{22}                                                                                                                                                                                                                                                                                                                                                                                                   & \begin{smallmatrix} r_{1'2} \sigma_{22} + \left( r_{21'} +r_{31'} \right)\sigma_{1'1'}  \\ + \imu \Omega_{1'3} \left(\sigma_{31'}^* - \sigma_{31'}\right) + r_{1'3} \sigma_{33} \end{smallmatrix} \\
      - r_{32} \sigma_{21'}                                                                                                                                                                                                                                 & \left(r_{32} + r_{1'2}\right) \sigma_{21'}                                                                                                                                                                                                                                                                                                                                                                                                     & -r_{1'2} \sigma_{21'}                                                                                                                                                                             \\
      \imu g a^* \sigma_{21'} + \left(r_{23} +r_{1'3} \right) \sigma_{31'}                                                                                                                                                                                  & -\imu g a^* \sigma_{21'} -  r_{23} \sigma_{31'}                                                                                                                                                                                                                                                                                                                                                                                                & -r_{1'3} \sigma_{31'}                                                                                                                                                                             \\
      -r_{32} \sigma_{23}                                                                                                                                                                                                                                   & \left( r_{32} + r_{1'2} \right) \sigma_{23}                                                                                                                                                                                                                                                                                                                                                                                                    & -r_{1'2} \sigma_{23}
    \end{matrix} \\[1em]
        &
    \left.
    \begin{matrix}
      0                                                                                                                                                & 0                                                                                                                                                       & 0                                                                                                                           \\
      0                                                                                                                                                & 0                                                                                                                                                       & 0                                                                                                                           \\
      \left( \gamma_{23} + \gamma_{1'2} - \gamma_{1'3}\right) \sigma_{31'}                                                                             & 0                                                                                                                                                       & \begin{smallmatrix} \left(2\gamma_{23} - r_{23} -r_{1'3} \right) \sigma_{33} {} & \\ + r_{32} \sigma_{22} \end{smallmatrix} \\
      0                                                                                                                                                & \begin{smallmatrix} \left(2\gamma_{1'3} - r_{2'1} - r_{31'} \right) \sigma_{1'1'} + {} & \\ r_{1'2} \sigma_{22} + r_{1'3} \sigma_{33} \end{smallmatrix} & 0                                                                                                                           \\
      \begin{smallmatrix} \left(2\gamma_{1'2} -r_{21'} -r_{31'}\right) \sigma_{1'1'} \\  + r_{1'2} \sigma_{22} + r_{1'3} \sigma_{33} \end{smallmatrix} & 0                                                                                                                                                       & \left(\gamma_{23} + \gamma_{1'2} - \gamma_{1'3}\right) \sigma_{31'}^*                                                       \\
      - r_{32} \sigma_{21'}                                                                                                                            & \imu g a^* \sigma_{21'}  + \left(r_{23} + r_{1'3} \right) \sigma_{31'}                                                                                  & -r_{32} \sigma_{23}                                                                                                         \\
      \left(r_{32} + r_{1'2}\right) \sigma_{21'}                                                                                                       & - \imu g a^* \sigma_{21'}  - r_{23} \sigma_{31'}                                                                                                        & \left( r_{32} + r_{1'2} \right) \sigma_{23}                                                                                 \\
      -r_{1'2} \sigma_{21'}                                                                                                                            & -r_{1'3} \sigma_{31'}                                                                                                                                   & -r_{1'2}\sigma_{23}                                                                                                         \\
      0                                                                                                                                                & 0                                                                                                                                                       & \imu g a \sigma_{21'}                                                                                                       \\
      0                                                                                                                                                & - 2 \imu \Omega_{1'3} \sigma_{31'}                                                                                                                      & - \imu g a \sigma_{31'}                                                                                                     \\
      \imu g a \sigma_{21'}                                                                                                                            & - \imu g a \sigma_{31'}                                                                                                                                 & 2 \imu g a \sigma_{23}
    \end{matrix}
    \right]
    \begin{bmatrix} a^* \\ a \\ \sigma_{23}^*  \\ \sigma_{31'}^* \\ \sigma_{21'}^* \\ \sigma_{33} \\ \sigma_{22} \\ \sigma_{1'1'}  \\ \sigma_{21'} \\ \sigma_{31'} \\ \sigma_{23} \end{bmatrix}\,.
    \nonumber
  \end{align}
\end{widetext}
\endgroup
\section{Quantum mechanical fluctuation terms within the generalized Maxwell-density matrix Langevin equations}
\label{sec:appendix-quant-fluc}
{The quantum-mechanical fluctuation terms for the three-level QCL quantum
  system are derived within the framework of the Langevin theory. In this paper, we have calculated the full diffusion matrix resulting from the c-number Langevin equations. Exploiting the positive semi-definiteness of the diffusion matrix, one can show that there exists a set of Ito stochastic differential equations equivalent to the Langevin equations. We can factorize the diffusion matrix to obtain a noise matrix that can be directly integrated into the Maxwell-density matrix approach for numerical modeling of fluctuations in dynamical optoelectronic devices. With a suitable choice of the noise matrix, one can guarantee a completely positive trace-preserving update map for long-term simulations. For the three-level QCL system, the fluctuation terms will fully account for the influence of the reservoirs and the properties of the nonlinear coupling between QCL system and optical field, including the incoherent tunneling transition, and can be represented as follows:}
\begingroup
\allowdisplaybreaks
\begin{widetext}
  \begin{subequations}
    \begin{align}
      \function{F_{23}}{t} ={}   & \function{\xi_{11}}{t} \sqrt{r_{32}}   +  \function{\xi_{14}}{t} \sqrt{r_{1'2}} - \function{\xi_{24}}{t} \sqrt{2\imu\mu_{z,23}  \function{E_z}{t} \function{\rho_{32}}{t} } - \function{\xi_{31}^*}{t} \frac{\imu \mu_{z,23} \function{E_z}{t}}{2}  + \function{\xi_{32}^*}{t} \frac{\imu \mu_{z,23} \function{E_z}{t}}{2}  \nonumber                                                            \\ & + \function{\xi_{33}^*}{t} \sqrt{\frac{\gamma_{1'2} - \gamma_{1'3} + \gamma_{23}}{2}} + \function{\xi_{41}^*} {t} \bigg[ \frac{\gamma_{1'3}-\gamma_{1'2}- \gamma_{23}}{2} + \frac{r_{32} \function{\rho_{22}}{t}}{2} + \frac{ 2 \gamma_{23} - r_{1'3} - r_{23}}{2} \function{\rho_{33}}{t} \nonumber \\ & - \frac{\mu_{z,23}^2 \function{E_z}{t}^2}{2} - r_{1'2} - r_{32} + \mu_{z,23} \function{E_z}{t} \abs{ \function{\rho_{32}}{t}}\bigg]^{1/2} \,,\\
      \function{F_{31'}}{t} ={}  & \function{\xi_{12}}{t} \sqrt{r_{1'3}} +  \function{\xi_{15a}}{t} \sqrt{r_{23}}     - \function{\xi_{15b}}{t}\imu \mu_{z,23} \function{E_z}{t} - \function{\xi_{25}}{t} \sqrt{-2 \imu\Omega_{31'} \function{\rho_{1'3}}{t}}  + \function{\xi_{31}}{t} \function{\rho_{1'3}}{t} \nonumber                                                                                                          \\ & + \function{\xi_{42}^*} {t}  \bigg[ \frac{2\gamma_{1'3} - \gamma_{1'2}-\gamma_{23}}{2} \function{\rho_{1'1'}}{t} + \frac{r_{1'2}}{2} \function{\rho_{22}}{t} - \abs{\function{\rho_{31'}}{t}}^2 + \frac{r_{1'3}}{2} \function{\rho_{33}}{t} - \mu_{z,23}^2 \function{E_z}{t}^2 - r_{1'3} - r_{23} \nonumber \\ & + \Omega_{3'1} \abs{\function{\rho_{31'}}{t}}\bigg]^{1/2} \,, \\
      \function{F_{21'}}{t} = {} & \function{\xi_{13}}{t} \sqrt{r_{1'2}} + \function{\xi_{16}}{t} \sqrt{r_{32}}        + \function{\xi_{32}}{t}  \function{\rho_{1'2}}{t}  +     \function{\xi_{33}^*}{t} \sqrt{\frac{\gamma_{1'2} - \gamma_{1'3} + \gamma_{23}}{2}} \function{\rho_{1'3}}{t} + \function{\xi_{43}^*}{t} \bigg[ \frac{2\gamma_{1'3} - \gamma_{1'2}-\gamma_{23}}{2} \nonumber                                        \\ & \times \function{\rho_{1'1'}}{t} + \frac{r_{1'2}}{2} \function{\rho_{22}}{t} - \abs{\function{\rho_{21'}}{t}}^2 + \frac{r_{1'3}}{2} \function{\rho_{33}}{t} - r_{1'2} - r_{32} - \frac{\gamma_{1'3} - \gamma_{1'2}-\gamma_{23}}{2} \abs{\function{\rho_{31'}}{t}}^2 \bigg]^{1/2} \,,\\
      \function{F_{33}}{t} = {}  & - \function{\xi_{11}^*}{t} \frac{\sqrt{r_{32}}\function{\rho_{32}}{t}}{2} - \function{\xi_{11}}{t} \frac{\sqrt{r_{32}}\function{\rho_{23}}{t}}{2} + \function{\xi_{12}^*}{t} \frac{\sqrt{r_{1'3}}\function{\rho_{1'3}}{t}}{2} + \function{\xi_{12}}{t} \frac{\sqrt{r_{1'3}}\function{\rho_{31'}}{t}}{2} + \function{\xi_{15a}^*}{t} \frac{\sqrt{r_{23}}\function{\rho_{1'3}}{t}}{2} \nonumber    \\ & + \function{\xi_{15a}}{t} \frac{\sqrt{r_{23}}\function{\rho_{31'}}{t}}{2} - \function{\xi_{15b}^*}{t} \frac{\function{\rho_{1'2}}{t}}{2} - \function{\xi_{15b}}{t} \frac{\function{\rho_{21'}}{t}}{2}  - \function{\xi_{16}^*}{t} \frac{\sqrt{r_{32}}\function{\rho_{1'2}}{t}}{2} - \function{\xi_{16}}{t} \frac{\sqrt{r_{32}}\function{\rho_{21'}}{t}}{2} \nonumber                     \\ & + \function{\xi_{21}}{t} \bigg\{  - \left(r_{32} + 1 \right) \abs{\function{\rho_{1'2}}{t}}^2 + r_{23} \big[\function{\rho_{33}}{t} - \abs{\function{\rho_{1'3}}{t}}^2 \big]   + r_{32} \big[ \function{\rho_{22}}{t} - \abs{\function{\rho_{32}}{t}}^2 \big] + \imu \mu_{z,23} \function{E_z}{t}  \big[\function{\rho_{32}}{t}\nonumber                     \\ & - \function{\rho_{23}}{t}\big] \bigg\}^{1/2} + \function{\xi_{22}}{t} \bigg\{ \imu \Omega_{31'} \big[-\function{\rho_{1'3}}{t} + \function{\rho_{31'}}{t}\big] + r_{31'}\function{\rho_{1'1'}}{t} + {r_{1'3}} \big[ \function{\rho_{33}}{t} - \abs{\function{\rho_{1'3}}{t}}^2 \big] \bigg\}^{1/2} \,,\\
      \function{F_{22}}{t} = {}  & \function{\xi_{11}^*}{t} \frac{\sqrt{r_{32}}\function{\rho_{32}}{t}}{2} + \function{\xi_{11}}{t} \frac{\sqrt{r_{32}}\function{\rho_{23}}{t}}{2}  + \function{\xi_{13}^*}{t} \frac{\sqrt{r_{1'2}}\function{\rho_{1'2}}{t}}{2} + \function{\xi_{13}}{t} \frac{\sqrt{r_{1'2}}\function{\rho_{21'}}{t}}{2} + \function{\xi_{14}^*}{t} \frac{\sqrt{r_{1'2}}\function{\rho_{32}}{t}}{2} \nonumber      \\ & + \function{\xi_{14}}{t} \frac{\sqrt{r_{1'2}}\function{\rho_{23}}{t}}{2} - \function{\xi_{15a}^*}{t} \frac{\sqrt{r_{23}}\function{\rho_{1'3}}{t}}{2} - \function{\xi_{15a}}{t} \frac{\sqrt{r_{23}}\function{\rho_{31'}}{t}}{2} + \function{\xi_{15b}^*}{t} \frac{\function{\rho_{1'2}}{t}}{2} + \function{\xi_{15b}}{t} \frac{\function{\rho_{21'}}{t}}{2} \nonumber \\   & + \function{\xi_{16}^*}{t} \frac{\sqrt{r_{32}}\function{\rho_{1'2}}{t}}{2} + \function{\xi_{16}}{t} \frac{\sqrt{r_{32}}\function{\rho_{21'}}{t}} {2}  - \function{\xi_{21}}{t} \bigg\{ r_{32} \function{\rho_{22}}{t} + r_{23} \function{\rho_{33}}{t} + \imu \mu_{z,23} \function{E_z}{t}  \big[\function{\rho_{32}}{t} - \function{\rho_{23}}{t}\big] \nonumber \\ & - \left({r_{32}} + 1 \right) \abs{\function{\rho_{1'2}}{t}}^2 - {r_{32}}\abs{\function{\rho_{32}}{t}}^2- {r_{23}} \abs{\function{\rho_{1'3}}{t}}^2 \bigg\}^{1/2} + \function{\xi_{23}}{t} \bigg\{r_{21'} \function{\rho_{1'1'}}{t} - {r_{1'2}}\big[\abs{\function{\rho_{1'2}}{t}}^2 -  \function{\rho_{22}}{t} \nonumber \\ & + \abs{\function{\rho_{23}}{t}}^2 \big]\bigg\}^{1/2} \,,\\
      \function{F_{1'1'}}{t} ={} & - \function{\xi_{12}^*}{t} \frac{\sqrt{r_{1'3}}\function{\rho_{1'3}}{t}}{2} - \function{\xi_{12}}{t} \frac{\sqrt{r_{1'3}}\function{\rho_{31'}}{t}}{2} - \function{\xi_{13}^*}{t} \frac{\sqrt{r_{1'2}}\function{\rho_{1'2}}{t}}{2} - \function{\xi_{13}}{t} \frac{\sqrt{r_{1'2}}\function{\rho_{21'}}{t}}{2} - \function{\xi_{14}^*}{t} \frac{\sqrt{r_{1'2}}\function{\rho_{32}}{t}}{2} \nonumber \\ & - \function{\xi_{14}}{t} \frac{\sqrt{r_{1'2}}\function{\rho_{23}}{t}}{2}  - \function{\xi_{22}}{t} \bigg\{ \imu \Omega_{31'} \big[-\function{\rho_{1'3}}{t} + \function{\rho_{31'}}{t}\big] + r_{31'}\function{\rho_{1'1'}}{t} + {r_{1'3}} \big[ \function{\rho_{33}}{t} - \abs{\function{\rho_{1'3}}{t}}^2 \big] \bigg\}^{1/2}  \nonumber \\ & - \function{\xi_{23}}{t} \bigg\{r_{21'} \function{\rho_{1'1'}}{t} - {r_{1'2}}\big[\abs{\function{\rho_{1'2}}{t}}^2 -  \function{\rho_{22}}{t} + \abs{\function{\rho_{23}}{t}}^2 \big]\bigg\}^{1/2}\,,\\
      \function{F_{1'2}}{t} ={}  & \function{\xi_{13}^*}{t} \sqrt{r_{1'2}}     + \function{\xi_{16}^*}{t} \sqrt{r_{32}}   + \function{\xi_{32}^*}{t}  \function{\rho_{21'}}{t} + \function{\xi_{33}}{t} \sqrt{\frac{\gamma_{1'2} - \gamma_{1'3} + \gamma_{23}}{2}} \function{\rho_{1'3}}{t}  + \function{\xi_{43}}{t} \bigg[ \frac{r_{1'2}}{2} \function{\rho_{22}}{t} \nonumber                                                    \\ & + \frac{2\gamma_{1'3} - \gamma_{1'2}-\gamma_{23}}{2} \function{\rho_{1'1'}}{t} - \abs{\function{\rho_{21'}}{t}}^2 + \frac{r_{1'3}}{2} \function{\rho_{33}}{t} - r_{1'2} - r_{32} - \frac{\gamma_{1'3} - \gamma_{1'2}-\gamma_{23}}{2} \abs{\function{\rho_{31'}}{t}}^2 \bigg]^{1/2} \,,\\
      \function{F_{1'3}}{t} ={}  & \function{\xi_{12}^*}{t} \sqrt{r_{1'3}} +\function{\xi_{15a}^*}{t} \sqrt{r_{23}}   + \function{\xi_{15b}^*}{t}\imu \mu_{z,23} \function{E_z}{t} + \function{\xi_{25}}{t} \sqrt{2 \imu\Omega_{31'} \function{\rho_{31'}}{t} } + \function{\xi_{31}^*}{t} \function{\rho_{31'}}{t} + \function{\xi_{42}} {t} \bigg[\frac{r_{1'2}}{2} \function{\rho_{22}}{t} \nonumber                             \\ & + \frac{2\gamma_{1'3} - \gamma_{1'2}-\gamma_{23}}{2} \function{\rho_{1'1'}}{t} - \abs{\function{\rho_{31'}}{t}}^2 + \frac{r_{1'3}}{2} \function{\rho_{33}}{t} - \mu_{z,23}^2 \function{E_z}{t}^2 - r_{1'3} - r_{23} + \Omega_{3'1} \abs{\function{\rho_{31'}}{t}}\bigg]^{1/2} \,,\\
      \function{F_{32}}{t} ={}   & \function{\xi_{11}^*}{t} \sqrt{r_{32}}   + \function{\xi_{14}^*}{t} \sqrt{r_{1'2}} + \function{\xi_{24}}{t} \sqrt{-2\imu\mu_{z,23}  \function{E_z}{t} \function{\rho_{23}}{t} } + \function{\xi_{31}}{t} \frac{\imu \mu_{z,23} \function{E_z}{t}}{2} - \function{\xi_{32}}{t} \frac{\imu \mu_{z,23} \function{E_z}{t}}{2} \nonumber                                                              \\ & + \function{\xi_{33}}{t} \sqrt{\frac{\gamma_{1'2} - \gamma_{1'3} + \gamma_{23}}{2}}  + \function{\xi_{41}} {t} \bigg[ \frac{\gamma_{1'3}-\gamma_{1'2}- \gamma_{23}}{2} + \frac{r_{32} \function{\rho_{22}}{t}}{2} + \frac{ 2 \gamma_{23} - r_{1'3} - r_{23}}{2} \function{\rho_{33}}{t} \nonumber \\ & - \frac{\mu_{z,23}^2 \function{E_z}{t}^2}{2}  - r_{1'2} - r_{32} + \mu_{z,23} \function{E_z}{t} \abs{ \function{\rho_{32}}{t}}\bigg]^{1/2}\,.
    \end{align}
  \end{subequations}
\end{widetext}
Here, the terms $\xi_{11}, \xi_{12}, \xi_{13}, \xi_{14}, \xi_{15a}, \xi_{15b}, \xi_{16},\xi_{31}, \xi_{32},\\ \xi_{33}, \xi_{41}, \xi_{42}, \xi_{43}$ are complex, while $\xi_{21}, \xi_{22}, \xi_{23}, \xi_{24}, \xi_{25}$ are real.
\endgroup
{
  \section{mbsolve simulation setups: THz HFC QCL setup}
  \label{sec:mboslve-python}
  Here, the Python script for setting up and running the mbsolve simulation of
  the THz HFC QCL setup is given. The script includes all input parameters for
  the description of the quantum system and the simulation scenario. The
  quantum mechanical description comprises the level occupations $\rho_{ii}$,
  the system Hamiltonian $\op{H_\mathrm{s}}$ with eigenenergies $E_i$ and
  anticrossing energies $\hbar\Omega_{ij}$, the dipole moment operator
  $\op{d}_z$, the scattering rates $r_{ij}$ and the dephasing rates
  $\gamma_{ij}$. For the one-dimensional dynamic Maxwell-density matrix
  simulations, the energy-resolved dephasing rates are simulated within the EMC
  approach and have to be averaged over the population inversions of the
  involved subbands, as it is described in detail in
  Refs.~\onlinecite{jirauschek2017density} and
  \onlinecite{jirauschek2017self}.}
\begin{widetext}
  \begin{lstlisting}[language=Python,caption={Code snippet of the Python script for the THz HFC QCL setup in  Ref.~\onlinecite{forrer2021self}.},label={lst:forrer2021}]

    import mbsolve.lib as mb
    import mbsolve.solvercpu
    import mbsolve.writerhdf5
    import mbsolve.readerhdf5
    
    import math
    import time
    
    # Hamiltonian
    energies = [ 0.0097 * mb.E0, 0.0082 * mb.E0, -0.0047 * mb.E0, 
    -0.0083 * mb.E0, -0.0097 * mb.E0 ]
    off_diagonales = [ 0.0005 * mb.E0, 0, 0, 0, 0, 0, 0, 0, 0, 0 ]
    H = mb.qm_operator(energies, off_diagonales)
    
    # dipole moment operator
    off_dipoles = [ 0, -2.9500e-09 * mb.E0, 0, 0, 0, 0, 0, 0, 0, 0 ]
    diag_dipoles = [ 0, 0, 0, 0, 0 ]
    u = mb.qm_operator(diag_dipoles, off_dipoles)
    
    # relaxation superoperator
    # scattering rate matrix R
    rates = [ [ 0, 1.8815e+09, 2.1290e+10, 4.0984e+09, 5.6000e+09 ],
              [ 3.5006e+09, 0, 3.2437e+08, 2.2854e+10, 2.0029e+12 ],
              [ 6.5578e+10, 6.2829e+08, 0, 8.0333e+11, 6.1577e+09 ],
              [ 6.8416e+09, 3.6845e+08, 6.6107e+11, 0, 4.7378e+12 ],
              [ 5.2192e+08, 6.7259e+10, 4.7554e+09, 4.7726e+12, 0 ] ]
    
    # pure dephasing rates
    pure_deph = [ 3.5857e+12, 9.3257e+11, 0, 0, 0, 0, 0, 0, 0, 0 ]
    relax_sop = mb.qm_lindblad_relaxation(rates, pure_deph)
    
    # initial density matrix 
    rho_init = mb.qm_operator([ 0.3705, 0.4937, 0.0741, 0.0333, 0.0285])
    
    h = 15e-6
    w = 60e-6
    A = h*w
    N = 6.35e21
    N_x = 2000
    l_device = 4e-3
    d_x = l_device / N_x
    f_noise = 1
    N_cell = N * A * d_x * f_noise
    
    qm = mb.qm_description(N, N_cell, H, u, relax_sop)
    loss = 760
    mat_ar = mb.material("AR_Forrer", qm, 12.96, 1, loss, 1.0)
    mb.material.add_to_library(mat_ar)
    
    dev = mb.device("5lvl")
    
    dev.add_region(mb.region("Active region", mat_ar, 0.0, l_device))
    
    # Scenario
    ic_d = mb.ic_density_const(rho_init)
    ic_e = mb.ic_field_const(0.0)
    sce = mb.scenario("hc_noise_forrer2021", N_x, 2200e-9,  ic_d, ic_e, ic_e)
    sce.add_record(mb.record("e1",0,4e-3))
    
    # run solver
    sol = mb.solver.create_instance("cpu-fdtd-5lvl-reg-cayley-qnoise", dev, sce)
    print('Solver ' + sol.get_name() + ' started')
    tic = time.time()
    sol.run()
    toc = time.time()
    print('Solver ' + sol.get_name() + ' finished in ' + str(toc - tic) + ' sec')
    
    # write results
    wri = mb.writer.create_instance("hdf5")
    outfile = dev.get_name() + "_" + sce.get_name() + "." + wri.get_extension()
    results = sol.get_results()
    wri.write(outfile, sol.get_results(), dev, sce)
    outfile_autosave = dev.get_name() + "_" + sce.get_name() + \
        "_autosave." + wri.get_extension()
    sim_data = sol.get_sim_data()
    wri.autosave(outfile_autosave, sim_data, dev, sce)
    
  \end{lstlisting}
\end{widetext}
\bibliographystyle{apsrev4-2}
\bibliography{references}

\end{document}